\documentclass[reprint]{revtex4-2}
\usepackage{amsmath,amsfonts,amscd,amssymb,amsthm}
\usepackage{commath}
\usepackage{bm}
\usepackage{dsfont}
\usepackage{graphicx}
\usepackage{epstopdf}
\usepackage{overpic}
\usepackage{cancel}
\usepackage{rotating}
\usepackage{url}
\usepackage{caption}
\usepackage{xcolor}
\usepackage{rotating}
\usepackage{multirow}
\usepackage{wrapfig}
\usepackage{mathtools}
\usepackage{subeqnarray}
\usepackage{setspace}
\usepackage{hyperref}
\usepackage{bbold}
\usepackage[group-separator={,}]{siunitx}

\usepackage[bottom,flushmargin,hang,multiple]{footmisc}
\usepackage{lipsum}

\begin{document} 

\title{\vspace{-.5in}\LARGE{The structure of global conservation laws \\ in  Galerkin plasma models}}

\author{Alan A. Kaptanoglu}
\affiliation{Department of Physics, University of Washington, Seattle, WA, 98195, USA} 
\author{Kyle D. Morgan}
\affiliation{Department of Aeronautics and Astronautics, University of Washington, Seattle, WA, 98195, USA}
\author{Chris J. Hansen}
\affiliation{Department of Aeronautics and Astronautics, University of Washington, Seattle, WA, 98195, USA}
\affiliation{Department of Applied Physics and Applied Mathematics, Columbia University, New York, NY, 10027, USA}
\author{Steven L. Brunton} 
\affiliation{Department of Mechanical Engineering, University of Washington, Seattle, WA, 98195, USA} 

\normalsize
\begin{abstract}{
Plasmas are highly nonlinear and multi-scale, motivating a hierarchy of models to understand and describe their behavior. However, there is a scarcity of plasma models of lower fidelity than magnetohydrodynamics (MHD). Galerkin models, obtained by projection of the MHD equations onto a truncated modal basis, can furnish this gap in the lower levels of the model hierarchy.
In the present work, we develop low-dimensional Galerkin plasma models which preserve global conservation laws by construction. This additional model structure enables physics-constrained machine learning algorithms that can discover these types of low-dimensional plasma models directly from data. 
This formulation relies on an energy-based inner product which takes into account all of the dynamic variables. 
The theoretical results here build a bridge to the extensive Galerkin literature in fluid mechanics, and facilitate the development of physics-constrained reduced-order models from plasma data.} 
%
\end{abstract}



\flushbottom
\maketitle

\section{Introduction}\label{Sec:introduction}
There are a tremendous number of known plasma models of varying model complexity, from magnetohydrodynamics (MHD) to the Klimontovich equations, but a large gap exists in the lower rungs of this hierarchy between simple circuit models and the many MHD variants. This is a valuable place for improvement because higher fidelity models often require computationally intensive and high-dimensional simulations~\cite{Candy2003,Ohia2012,Groselj2018}, obfuscating the dynamics and precluding model-based real-time control. Additionally, many high-dimensional nonlinear systems tend to evolve on low-dimensional attractors~\cite{Taira2017aiaa}, defined by spatio-temporal coherent structures that characterize the dominant behavior of the system. 
A number of studies in the plasma physics community indicate that the vast majority of the total plasma energy can be explained by fewer than ten low-dimensional modes, across a large range of parameter regimes, geometry, and degree of nonlinearity~\cite{furno2001fast,jimenez2007analysis,vanMilligen14,pandya,victor2015development,strait2016spatial,byrne2017study,gu2019new,kaptanoglu2020}. 
In these cases, the evolution of only a few coherent structures, obtained from systematic model-reduction techniques~\cite{benner2005dimension, benner2017model}, can closely approximate the full evolution of the high-dimensional physical system.

In the present work and our companion paper~\cite{kaptanoglu2020physics}, we provide a theoretical framework for physics-constrained, low-dimensional plasma models which furnish this important gap in the hierarchy of plasma models. This is a breakthrough in principled and interpretable model reduction, offering an alternative to deep learning methods which often require vast quantities of data and produce opaque models. The present work focuses on the important technical details for constructing and constraining these models, while the accompanying publication contains an overview of our high-level contributions and initial results on 3D plasma simulations. 
\section{Low-dimensional models}\label{Sec:pod}
Although there are many ways to obtain low-dimensional models, Galerkin methods and their extensions have seen remarkable success in fluid mechanics; careful development of a dimensionalized inner product enabled the extension of the proper orthogonal decomposition (POD) from incompressible to compressible fluid flows~\cite{rowley2004model}. 
It is also common in fluid mechanics to obtain nonlinear reduced-order models by Galerkin projection of the Navier-Stokes equations onto POD modes, making it possible to enforce known symmetries and conservation laws, such as conservation of energy~\cite{noack2011galerkin,balajewicz2013low,Schlegel2015jfm,Carlberg2017jcp}. 
%

The present work adapts and extends these innovations for plasmas, enabling a wealth of advanced modeling and control machinery. The POD is already used extensively for interpreting plasma physics data across a range of parameter regimes~\cite{dudok1994biorthogonal,levesque2013multimode,galperti2014development,van2014use,hansen2015numerical}. For clarity of presentation and robust connection with the Galerkin literature in fluid mechanics, we present results for MHD models which are at most quadratic in nonlinearity. This includes ideal MHD, incompressible Hall-MHD, or compressible Hall-MHD with a slowly time-varying density, which together describe the dynamics of a fairly broad class of space and laboratory plasmas~\cite{Schnack2006,Ma2001,Krishan2004,Ebrahimi2011,Ferraro2012,kaptanoglu2020two}.

\subsection{An MHD energy inner product\label{energy_inner_prod}}
Traditional use of the POD on the MHD fields (velocity, magnetic, and temperature) requires separate decompositions for $\bm{u}$, $\bm{B}$, and $T$, or an arbitrary choice of dimensionalization. However, separate decompositions of the fields obfuscates the interpretability and increases the complexity of a low-dimensional model, and choosing the units of the combined matrix of measurement data can have a significant impact on the performance and energy spectrum of the resulting POD basis. %
Inspired by the inner product defined for compressible fluids~\cite{rowley2004model}, we introduce an inner product for compressible magnetohydrodynamic fluids by using the configuration vector
$\bm{q}(\bm{x},t)= [\bm{B}_u,\bm{B},B_T]$. Here 
\begin{align}
\label{eq:rescale}
\bm{B}_u = \sqrt{\rho\mu_0}\bm{u}, && B_T = 2\sqrt{{\rho \mu_0k_bT}/{m_i(\gamma-1)}},
\end{align}
where $\bm{u}$ is the fluid velocity, $\rho$ is the mass density, $k_b$ is Boltzmann's constant, $\mu_0$ is the permeability of free space, $T$ is the plasma temperature, $m_i$ is the ion mass, $\gamma$ is the adiabatic index, and ${p = 2\rho T/m_i}$ is the plasma pressure. 
$\bm{B}_u$ and $B_T$ are defined so that the following scaled inner product yields the total energy $W,$ 
\begin{equation}
    \label{eq:inner_prod}
W = \frac{1}{2\mu_0}\langle\bm{q},\bm{q}\rangle = \int \left(\frac{1}{2}\rho u^2+\frac{B^2}{2\mu_0}+\frac{p}{\gamma-1}\right) d^3\bm{x}.
\end{equation}
\subsection{Review of the POD\label{sec:pod_review}}
For the POD, measurements at time $t_k$ are arranged in a vector $\bm{q}_k\in\mathds{R}^D$, called a snapshot, where the dimension $D$ is the product of the number of spatial locations and the number of variables measured at each point.  
The data is sampled at times $t_1,\hspace{0.05in}t_2,\hspace{0.05in}...,\hspace{0.05in}t_M$, arranged in a matrix  $\bm{X}\in\mathds{R}^{D\times M}$, and the average in time $\bar{\bm{q}}$ is subtracted off. The singular value decomposition (SVD) provides a low-rank approximation of the subsequent data matrix
 \medmuskip=2mu
 \thinmuskip=2mu
 \thickmuskip=2mu
\begin{eqnarray}
\bm{X}
= \overset{\text{\normalsize time}}{\left.\overrightarrow{\overset{~~}{\begin{bmatrix}
q_1(t_1) & q_1(t_2) & \cdots & q_1(t_M)\\
q_2(t_1) & q_2(t_2) & \cdots & q_2(t_M)\\
\vdots & \vdots & \ddots & \vdots \\
q_D(t_1) & q_D(t_2) & \cdots & q_D(t_M)
\end{bmatrix}}}\right\downarrow}\begin{rotate}{270}\hspace{-.125in}state~~\end{rotate} \hspace{.125in}= \bm{U}\bm{\Sigma}\bm{V}^*,
\label{Eq:DataMatrix}
\end{eqnarray}
where $\bm{U}\in\mathds{R}^{D\times D}$ and $\bm{V}\in\mathds{R}^{M\times M}$ are unitary matrices, and $\bm{\Sigma} \in \mathds{R}^{D\times M}$ is a diagonal matrix containing non-negative and decreasing entries $s_{jj}$ called the singular values of $\bm{X}$. $\bm{V}^*$ denotes the complex-conjugate transpose of $\bm{V}$.  
The singular values indicate the relative importance of the corresponding columns of $\bm{U}$ and $\bm{V}$ for describing the spatio-temporal structure of $\bm{X}$. Although varying terminology is used in different fields (in the plasma physics community this method is often referred to as the biorthogonal decomposition, or BOD), in practice the SVD, BOD, and POD are synonymous.
It is often possible to discard small values of $\bm{\Sigma}$, resulting in a truncated matrix $\bm{\Sigma}_r\in\mathds{R}^{r\times r}$. With the first $r\ll \min(D,M)$ columns of $\bm{U}$ and $\bm{V}$, denoted $\bm{U}_r$ and $\bm{V}_r$, the matrix $\bm{X}$ can be approximated as
\begin{equation}
\bm{X}\approx \bm{U}_r\bm{\Sigma}_r\bm{V}_r^*.
\label{eq:svd_trunc}
\end{equation}
The truncation rank $r$ is typically chosen to balance accuracy and complexity~\citep{brunton2019data}. The computational complexity of the SVD is $\mathcal{O}(DM^2+M^3)$~\cite{golub1996cf}, although there are randomized singular value decompositions~\cite{frieze2004fast,liberty2007randomized,woolfe2008fast} for extremely large problems that can be as fast as $\mathcal{O}(DM\log(r))$. This computational speed enables the SVD to be computed online, for updating a model in a real-time control application, or offline, for the decomposition of very large data, an examination of the physics, or development of a more generic model describing the dynamics of an amalgam of discharges.
A well-defined SVD requires that the measurements in $\bm{X}$ have the same physical dimensions.
%
With a dimensionalized measurement vector $\bm{q}$, the matrix $\bm{X}^*\bm{X}$ may be computed via
\medmuskip = 0.5mu 
\begin{align}
    \label{eq:snapshots}
    \bm{X}^*\bm{X} \approx 
    \langle\bm{q}(t_k),\bm{q}(t_m)\rangle. 
\end{align}
When the number of snapshots is far fewer than the number of measurements, $M\ll D$, we can use the method of snapshots. Substitution of Eq.~\ref{eq:svd_trunc} into $\bm{X}^*\bm{X}$ produces
\begin{equation}    
    \label{eq:snapshots}
    \bm{X}^*\bm{X}\bm{V}_r \approx 
    \bm{V}_r\bm{\Sigma}_r^2,
\end{equation}
an eigenvalue equation for $\bm{V}_r$; therefore we can obtain $\bm{V}_r$ by diagonalizing $\bm{X}^*\bm{X}\in\mathds{R}^{M\times M}$ instead of computing the SVD of $\bm{X}$. The \textit{chronos} are the temporal SVD modes, i.e. the columns of $\bm{V}_r$, denoted $\bm{v}_{j}$. The \textit{topos} are the spatial modes forming the columns of $\bm{U}_r$, denoted $\bm{\chi}$. 
In the present work, we scale the normalized matrix of chronos, $\bm{a}$, through
$    a_{jk} = 
    {v_{jk}}/{\sum_{j=1}^r\max_k|v_{jk}|}$.
Finally, the reconstruction can be written
\begin{equation}
    \label{eq:q_expansion}
    \bm{q}(\bm{x}_i,t_k) \approx \bar{\bm{q}}(\bm{x}_i) + 
    \sum_{j=1}^r \bm{\chi}_j(\bm{x}_i)a_j(t_k),
\end{equation}
We have absorbed the normalization of $a_{jk}$ and the singular values into the definition of $\bm{\chi}_j(\bm{x}_i)$. By construction $\langle \bm{\chi}_i,\bm{\chi}_j \rangle \propto \delta_{ij}$. 
Note that, in principle, we could have expanded $\bm{q}$ in any set of modes, although orthonormal modes are preferred because this property facilitates the analysis in Section~\ref{sec:conservation_laws}. 
The advantage of the POD basis is that the modes are ordered by energy content; a truncation of the system still captures the vast majority of the dynamics. 
A separate POD of each of the MHD fields would lead to three sets of POD modes with independent time dynamics and mixed orthogonality properties.
In contrast, our approach captures all the fields simultaneously, resulting in a single set of modes $a_i(t)$ in Eq.~\eqref{eq:q_expansion}. 

\begin{figure*}
    \centering
    \begin{overpic}[width=0.95\linewidth]{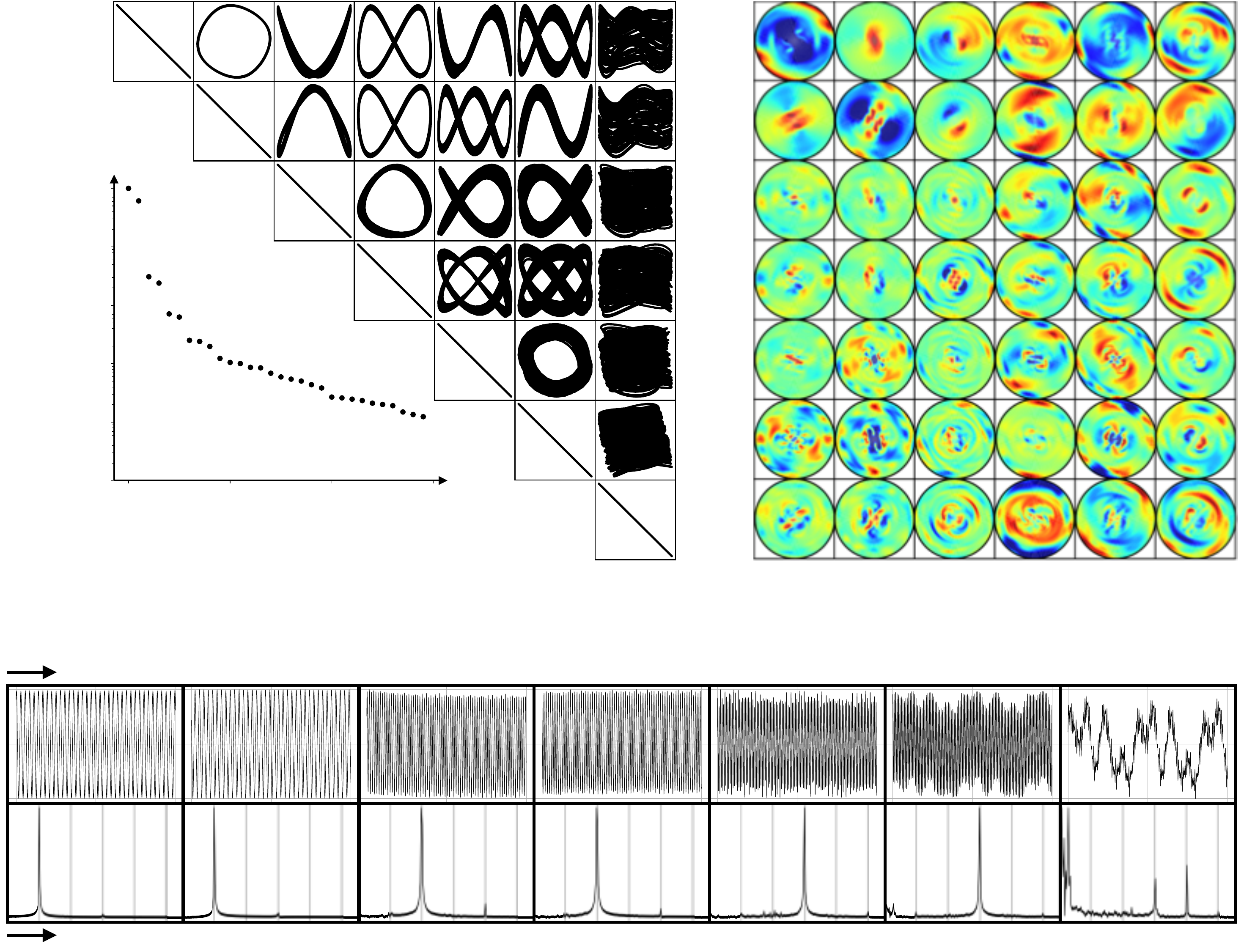}
    \put(1.75,23.5){$t$}
    \put(1.25,-0.75){$\omega$}
    \put(-1,81.5){(a) Pairwise correlations of POD amplitudes}
    \put(59,81.5){(b) Spatial POD modes}
    \put(-1,26.75){(c) Time evolutions $a(t)$ and Fourier transforms $\tilde{a}(\omega)$}
    \put(2.5,39){\begin{turn}{90}\footnotesize{Normalized singular values}\end{turn}}
    \put(18,33.5){\footnotesize{Mode index}}
    \footnotesize{
    \put(5.75,61){$10^0$}
    \put(5.25,56.25){$10^{-1}$}
    \put(5.25,51.5){$10^{-2}$}
    \put(5.25,46.75){$10^{-3}$}
    \put(5.25,42){$10^{-4}$}
    \put(5.25,37.5){$10^{-5}$}
    \put(9.9,36){$0$}
    \put(17.5,36){$10$}
    \put(25.75,36){$20$}
    \put(34,36){$30$}
    }
    \normalsize{
    \put(6.25,73){$a_1$}
    \put(12.5,66.5){$a_2$}
    \put(19,60){$a_3$}
    \put(25.5,53.5){$a_4$}
    \put(32,47.25){$a_5$}
    \put(38.5,41){$a_6$}
    \put(45,34.25){$a_7$}
    \put(11.5,77.75){$a_1$}
    \put(18,77.75){$a_2$}
    \put(24.5,77.75){$a_3$}
    \put(30.75,77.75){$a_4$}
    \put(37.25,77.75){$a_5$}
    \put(43.5,77.75){$a_6$}
    \put(50,77.75){$a_7$}
    \put(57.5,73){$\chi_1$}
    \put(57.5,66.5){$\chi_2$}
    \put(57.5,60){$\chi_3$}
    \put(57.5,53.5){$\chi_4$}
    \put(57.5,47.25){$\chi_5$}
    \put(57.5,41){$\chi_6$}
    \put(57.5,34.5){$\chi_7$}
    \put(62.75,77.75){$B_x$}
    \put(69.25,77.75){$B_y$}
    \put(75.75,77.75){$B_z$}
    \put(82.1,77.75){$B_x^u$}
    \put(88.5,77.75){$B_y^u$}
    \put(94.9,77.75){$B_z^u$}
    }
    \scriptsize{
    \put(-0.5,20.5){1}
    \put(-0.5,16.25){0}
    \put(-1,12.35){-1}
    \put(-0.5,10.85){1}
    \put(-0.5,2){0}
    }
    \put(2.65,7){\small{$1$}}
    \put(5.25,7){\small{$2$}}
    \put(7.75,7){\small{$3$}}
    \put(10.25,7){\small{$4$}}
    \put(12.75,7){\small{$5$}}
    \put(7,22.5){\normalsize{$a_1$}}
    \put(21,22.5){\normalsize{$a_2$}}
    \put(35,22.5){\normalsize{$a_3$}}
    \put(49,22.5){\normalsize{$a_4$}}
    \put(63,22.5){\normalsize{$a_5$}}
    \put(77,22.5){\normalsize{$a_6$}}
    \put(91,22.5){\normalsize{$a_7$}}
    \put(7,0.25){\normalsize{$\tilde{a}_1$}}
    \put(21,0.25){\normalsize{$\tilde{a}_2$}}
    \put(35,0.25){\normalsize{$\tilde{a}_3$}}
    \put(49,0.25){\normalsize{$\tilde{a}_4$}}
    \put(63,0.25){\normalsize{$\tilde{a}_5$}}
    \put(77,0.25){\normalsize{$\tilde{a}_6$}}
    \put(91,0.25){\normalsize{$\tilde{a}_7$}}
    \end{overpic}
    \caption{The first seven POD modes for an isothermal, compressible Hall-MHD simulation of the HIT-SI device~\cite{jarboe2006spheromak}: (a) Feature space trajectories of every mode pair and the singular values; (b) \textcolor{black}{3D spatial modes visualized in the $Z = 0$ midplane and normalized to $\pm 1$}; (c) Normalized temporal modes and corresponding Fourier transforms produce harmonics of the driving frequency, labeled 1-5 in the Fourier space.}
    \label{fig:pod_modes}
\end{figure*}

An example of this decomposition is illustrated in Fig.~\ref{fig:pod_modes} for a 3D plasma simulation that is dominated by harmonics of a driving frequency. In general, examining the structure and symmetry in the spatial and temporal POD modes can inform physical understanding. For instance, in Fig.~\ref{fig:pod_modes}, the short-wavelength structures exhibited in the 3D spatial modes derive both from dispersive whistler waves via the Hall term and the small characteristic scale associated with the driving actuator. 
The steep fall-off in the singular values also indicates that models of only the first few modes would be enough to accurately forecast and control the dominant dynamics. 

%
%

\subsection{\label{sec:generic_pod_galerkin} POD-Galerkin models}
While we have obtained a useful modal decomposition of the evolved fields, we have yet to derive a model for the subsequent temporal evolution of the modes. 
Now we substitute Eq.~\eqref{eq:q_expansion} into a quadratically nonlinear MHD model, such as ideal MHD, incompressible Hall-MHD, or compressible Hall-MHD with a slowly time-varying density. Utilizing the orthonormality of the $\bm{\chi}_i$ produces: 
\begin{align}
\label{eq:Galerkin_model}
\dot{a}_i(t) &= C_i^0 + \sum_{j=1}^rL^0_{ij}a_j + \sum_{j,k=1}^r Q^0_{ijk}a_ja_k, \\ \notag
C_i^0 &= \langle \bm{C} + \bm{L}(\bar{\bm{q}}) + \bm{Q}(\bar{\bm{q}},\bar{\bm{q}}),\bm{\chi}_i\rangle, \\ \notag
L^0_{ij} &= \langle \bm{L}(\bm{\chi}_j) + \bm{Q}(\bar{\bm{q}},\bm{\chi}_j) + \bm{Q}(\bm{\chi}_j,\bar{\bm{q}}),\bm{\chi}_i\rangle, \\ \notag
Q^0_{ijk} &= \langle \bm{Q}(\bm{\chi}_j,\bm{\chi}_k),\bm{\chi}_i\rangle.
\end{align}
The inner products integrate out the spatial dependence, and the model is quadratic in the temporal POD modes $a_i(t)$. 
In contrast to Eq.~\eqref{eq:Galerkin_model}, a Galerkin model based on separate POD expansions for each field would involve significantly more complicated nonlinear terms from mixing and a lack of orthonormality $\langle \bm{\chi}^{\bm{u}}_i,\bm{\chi}^{\bm{B}}_j \rangle \neq \delta_{ij}$ between the different POD modes associated with each field. 
\section{Constraints on model structure\label{sec:conservation_laws}}
Local and global MHD conservation laws are retained in this low-dimensional basis.
Vanishing $\nabla\cdot\bm{B}$ and the orthonormality of the temporal POD modes produce
\begin{align}
    \nabla\cdot\bm{\chi}_i^B = 0,\,\,\,\,\,\,\forall i.
\end{align}
In other words, the orthonormality of the temporal modes guarantees that the local divergence constraint is satisfied by each of the $\bm{\chi}_i^B$ by construction. 
%
In contrast, global energy conservation will produce strong constraints on the structure of the coefficients in Eq.~\ref{eq:Galerkin_model}.
\subsection{\label{sec:global_energy}Global conservation of energy}
For an examination of the global conservation laws, we consider isothermal Hall-MHD with the assumption that the density is slowly-varying in time. This model reduces to ideal MHD and incompressible resistive or Hall MHD in the appropriate limits, and produces (Galtier~\cite{galtier2016introduction} Eq. 3.22)
\medmuskip=-2mu
\thickmuskip=-1mu
\thinmuskip=-1mu
\begin{align}
\label{eq:galtier_deriv}
    \frac{\partial W}{\partial t} = &- \int \left[\tilde{\nu}(\nabla\times\bm{u})^2+\frac{\eta}{\mu_0}(\nabla\times\bm{B})^2 + \frac{4}{3}\tilde{\nu}(\nabla\cdot\bm{u})^2\right] d^3\bm{x}  \\ \notag
    &-\oint \left[\left(\frac{1}{2}\rho u^2+p\right)\bm{u} + \bm{P} - \frac{4}{3}\tilde{\nu}(\nabla\cdot\bm{u})\bm{u} - \tilde{\nu}\bm{u}\times(\nabla\times\bm{u})\right]\cdot\hat{\bm{n}}dS.
\end{align}
Here $\hat{\bm{n}}$ is a unit normal vector to the boundary, and 
\medmuskip=1mu
\thickmuskip=1mu
\thinmuskip=1mu
\begin{align}
\label{eq:poyntingvec}
\bm{P} = \frac{1}{\mu_0}\bm{E}\times\bm{B} = \frac{\bm{u}_e}{\mu_0}\cdot(B^2\bm{I} - \bm{B}\bm{B})  - \frac{\eta}{\mu_0^2}(\nabla\times\bm{B})\times\bm{B},
\end{align}
is the Poynting vector ($\bm{E}$ is the electric field), which is often a
known function of space and time. Omission of the Hall term changes $\bm{u}_e$ to $\bm{u}$. Even with a Hall-MHD model that evolves the temperature, the electron diamagnetic contribution to $\bm{P}$ does not change the energy balance if Dirichlet boundary conditions are used for $\rho$ and $T$. To simplify, we assume that $\bm{u}\cdot\hat{\bm{n}} = \bm{u}\times\hat{\bm{n}} = 0$, $\bm{J}\cdot\hat{\bm{n}} = 0$, and $\bm{B}\cdot\hat{\bm{n}} = 0$ at the wall, consistent with the simulations used in the accompanying work~\cite{kaptanoglu2020physics}.
Moreover, we now assume that we have a steady-state, define a constant $a_0(t) = 1$, and substitute Eq.~\eqref{eq:q_expansion} into Eq.~\eqref{eq:galtier_deriv}. 
\medmuskip=1mu
\begin{widetext}
\begin{align}
\label{eq:W_decomp_reduced}
    0 \approx \frac{\partial W}{\partial t} &= \oint\frac{\eta}{\mu_0^2} ((\nabla\times\bm{B})\times\bm{B})\cdot\hat{\bm{n}}dS - \int \left[\frac{\nu}{\mu_0}(\nabla\times\bm{B}_u-\frac{\nabla\rho}{2\rho}\times\bm{B}_u)^2+\frac{\eta}{\mu_0^2}(\nabla\times\bm{B})^2 + \frac{4}{3}\frac{\nu}{\mu_0}(\nabla\cdot\bm{B}_u-\frac{\nabla\rho}{2\rho}\cdot\bm{B}_u)^2\right] d^3\bm{x}, \\
\notag
&= W^C + \sum_{i=1}^rW_i^L a_i + \sum_{i,j=1}^rW_{ij}^\text{Q}a_ia_j = W^C+\sum_{i=1}^ra_i(W_i^L + \sum_{j=1}^rW_{ij}^Qa_j) 
= W^C+\sum_{i=1}^r\sum_{j=0}^rW_{ij}^Qa_ia_j = \sum_{i=0}^r\sum_{j=0}^rW_{ij}^Qa_ia_j,
\end{align}
where we have padded the matrix in the last step so that $W_{0i}^Q = 0$, $W_{i0}^Q = W_i^L$ for $i \in \{1,...,r\}$, and $W^Q_{00} = W^C$. It immediately follows from Eq.~\eqref{eq:W_decomp_reduced}
that $W^Q_{ij}$ is an antisymmetric matrix. Writing out the coefficients we have
\medmuskip=-2mu
\thinmuskip=-2mu
\thickmuskip=-2mu
\begin{align}
0 = W^\text{Q}_{00} = &\frac{\eta}{\mu_0}\oint \left[(\nabla\times\bar{\bm{B}})\times\bar{\bm{B}}\right]\cdot\hat{\bm{n}}dS -\int\left[\nu(\nabla\times\bar{\bm{B}}_u-\frac{\nabla\rho}{2\rho}\times\bar{\bm{B}}_u)^2+ \frac{\eta}{\mu_0}(\nabla\times\bar{\bm{B}})^2+\frac{4}{3}\nu(\nabla\cdot\bar{\bm{B}}_u-\frac{\nabla\rho}{2\rho}\cdot\bar{\bm{B}}_u)^2\right]d^3\bm{x}, \\ \notag
0 = W_{i0}^\text{Q} = &\frac{\eta}{\mu_0}\oint \left[(\nabla\times\bar{\bm{B}})\times\bm{\chi}_i^{B} +(\nabla\times\bm{\chi}_i^B)\times\bar{\bm{B}} \right]\cdot\hat{\bm{n}}dS \\ \notag &-2\int\left[\nu(\nabla\times\bar{\bm{B}}_u-\frac{\nabla\rho}{2\rho}\times\bar{\bm{B}}_u)\cdot(\nabla\times\bm{\chi}^{B_u}_i-\frac{\nabla\rho}{2\rho}\times\bm{\chi}^{B_u}_i)+\frac{\eta}{\mu_0}(\nabla\times\bar{\bm{B}})\cdot(\nabla\times\bm{\chi}^{B}_i)+ \frac{4}{3}\nu(\nabla\cdot\bar{\bm{B}}_u-\frac{\nabla\rho}{2\rho}\cdot\bar{\bm{B}}_u)\cdot(\nabla\cdot\bm{\chi}^{B_u}_i-\frac{\nabla\rho}{2\rho}\cdot\bm{\chi}^{B_u}_i) \vphantom{\nu(\nabla\times\bm{\chi}^{B_u}_i-\frac{\nabla\rho}{2\rho}\times\bm{\chi}^{B_u}_i)\cdot(\nabla\times\bm{\chi}^{B_u}_j}\right] d^3\bm{x}, \\ \notag
W_{ij}^\text{Q} = &\frac{\eta}{\mu_0}\oint \left[(\nabla\times\bm{\chi}_i^B)\times\bm{\chi}_j^B\right]\cdot\hat{\bm{n}}dS \\ \notag &-\int\left[\nu(\nabla\times\bm{\chi}^{B_u}_i-\frac{\nabla\rho}{2\rho}\times\bm{\chi}^{B_u}_i)\cdot(\nabla\times\bm{\chi}^{B_u}_j-\frac{\nabla\rho}{2\rho}\times\bm{\chi}^{B_u}_j)+\frac{\eta}{\mu_0}(\nabla\times\bm{\chi}^{B}_i)\cdot(\nabla\times\bm{\chi}^{B}_j)+  \frac{4}{3}\nu(\nabla\cdot\bm{\chi}^{B_u}_i-\frac{\nabla\rho}{2\rho}\cdot\bm{\chi}^{B_u}_i)\cdot(\nabla\cdot\bm{\chi}^{B_u}_j-\frac{\nabla\rho}{2\rho}\cdot\bm{\chi}^{B_u}_j)\vphantom{\nu(\nabla\times\bm{\chi}^{B_u}_i-\frac{\nabla\rho}{2\rho}\times\bm{\chi}^{B_u}_i)\cdot(\nabla\times\bm{\chi}^{B_u}_j}\right]d^3\bm{x}. 
\end{align}
\medmuskip=-1mu
\thickmuskip=2mu
\thinmuskip=-1mu
\end{widetext}
With some algebra, we can compute $a_i\dot{a}_i$ for $i \in \{1,...,r\}$,
\begin{align}
\label{eq:q2evo}
    a_i\dot{a}_i = \sum_{i,j=1}^ra_i\frac{\partial a_j}{\partial t}\int \chi_i \chi_j d^3\bm{x} = \int \frac{1}{2}\frac{\partial q^2}{\partial t} d^3\bm{x} = \frac{\partial W}{\partial t} 
\end{align}
In index notation $a_i\dot{a}_i = a_iC^0_i +  a_iL^0_{ij}a_j + a_i Q^0_{ijk}a_ja_k$ for $i,j,k \in \{1,...,r\}$. 
First, note that $W^\text{Q}_{i0}=0$ produces $C^0_i = 0$ for all $i \in \{1,...,r\}$. In other words, there are no constant terms in the Galerkin model; data-driven methods can implement this constraint by simply searching for models that do not have constant terms. This a physical consequence of our assumption that $\bar{q}$ is steady-state because nonzero constant terms in the Galerkin model would imply the possibility of unbounded growth in the energy norm. The anti-symmetry of $W_{ij}^Q$ for $i,j \in \{1,...,r\}$ constrains the quadratic structure of $\bm{a}^T\cdot\bm{a}$, 
\begin{align}
\label{eq:L_constraint}
    \bm{a}^T\bm{L}^0\bm{a} \approx 0.
\end{align}
This physical interpretation is also clear; if the plasma is steady-state but has finite dissipation, the input power, here manifested through a purely quadratic Poynting flux $\bm{P} \propto \eta\bm{J}\times\bm{B}$, must be balancing these losses.  Finally, because of the boundary conditions there are no cubic terms in the energy at all, meaning 
\begin{align}
\label{eq:Q_constraint}
    \bm{a}^T\bm{Q}^0\bm{a}\bm{a} = 0. 
\end{align}
This is the condition for a system to have energy-preserving quadratic nonlinearities; this conclusion does not rely on any assumption of steady-state and energy-preserving structure in other quadratic nonlinearities is well-studied in fluid mechanics~\cite{Schlegel2015jfm,Loiseau2018jfm}.
\subsection{\label{sec:global_helicity}Global conservation of cross-helicity}
An analogous derivation can be done to further constrain the model-building for systems which conserve cross-helicity, although this is not appropriate for the simulation results in the accompanying work~\cite{kaptanoglu2020physics}. 
Consider the local form of cross-helicity $H_c = \bm{u}\cdot\bm{B}$. Using Galtier~\cite{galtier2016introduction} Eq. (3.36), 
\begin{widetext}
\begin{align}
    \frac{\partial H_c}{\partial t} = &-\nabla\cdot\left[\left(\frac{u^2}{2}+ \frac{\gamma p}{(\gamma-1)\rho}\right)\bm{B} + \bm{u}\times(\bm{u}\times\bm{B})-\frac{d_i}{\sqrt{\rho\mu_0}}\bm{u}\times\left((\nabla\times\bm{B})\times\bm{B}\right) - \eta\bm{u}\times(\nabla\times\bm{B})  \right] \\ \notag &+ \nu \nabla\cdot\left(\bm{B}\times\bm(\nabla\times\bm{u}) + \frac{4}{3}(\nabla\cdot\bm{u})\bm{B}\right)-\frac{d_i}{\sqrt{\rho\mu_0}}(\nabla\times\bm{u})\cdot\left((\nabla\times\bm{B})\times\bm{B}\right) - (\eta+\nu)(\nabla\times\bm{B})\cdot(\nabla\times\bm{u}).
\end{align}
Consider again the simplifying case $\bm{J}\cdot\hat{\bm{n}} = 0$, $\bm{B}\cdot\hat{\bm{n}} = 0$, and $\bm{u}\cdot\hat{\bm{n}} = \bm{u}\times\hat{\bm{n}} = 0$. Then the integral form is
\medmuskip=0mu
\begin{align}
    0 \approx \int \frac{\partial H_c}{\partial t}d^3\bm{x} = \int \left[\nu \frac{\nabla\rho}{\rho}\cdot\left(\bm{B}\times(\nabla\times\bm{u})+\frac{4}{3}(\nabla\cdot\bm{u})\bm{B}\right)-\frac{d_i}{\sqrt{\rho\mu_0}}(\nabla\times\bm{u})\cdot\left((\nabla\times\bm{B})\times\bm{B}\right) - (\eta+\nu)(\nabla\times\bm{B})\cdot(\nabla\times\bm{u}) \right]d^3\bm{x}.
\end{align}
\end{widetext}
Substituting in Eq.~\eqref{eq:q_expansion} produces terms up to cubic in the temporal POD modes,
\begin{align}
\label{eq:Hc_constraints}
    \int\frac{\partial H_c}{\partial t}d^3\bm{x} &= \frac{\partial}{\partial t}(a_ia_j)\int \frac{1}{\sqrt{\rho\mu_0}}\bm{\chi}_i^{B_u}\cdot\bm{\chi}_j^Bd^3\bm{x}  \\ \notag 
    0 &= A_{ij}\frac{\partial}{\partial t}(a_ia_j) = 
    \begin{cases}
    A_{ij}C^0_ja_i \\
    A_{ij}L^0_{jk}a_ia_k \\
    A_{ij}Q^0_{jkl}a_ia_ka_l 
    \end{cases}
\end{align}
Note that if the system is energy-preserving, $C^0_j = 0$ for all j, so the first equality is already satisfied. The second inequality gives $A_{ij}L^0_{jk}$ anti-symmetric under swapping $i$ and $k$, and energy-preservation gave anti-symmetry under swapping $j$ and $k$ (Eq. \ref{eq:L_constraint}). The most straightforward solution is $L^0_{jk} = 0$ for all $j$,$k$;
this solution is precisely the ideal limit corresponding to $\eta = \nu = 0$. Since $A_{ij}$ is not symmetric by construction, this constraint can also apply to systems which conserve cross-helicity despite finite dissipation.

Lastly, $A_{ij}Q^0_{jkl}$, containing only the contribution from the Hall-term, exhibits the same structure as (and is compatible with) our constraint on the energy-preserving nonlinearities in Eq.~\eqref{eq:Q_constraint}. The simplest solution is $A_{ij}Q^0_{jkl}=0$ for all $i,k,l$, since this corresponds to standard MHD without the Hall term. Like the analysis of the linear terms, this constraint indicates that if the Hall-terms have this special energy-preserving structure, nonzero Hall contributions can still conserve cross-helicity. 
%
%
Enforcing other invariants of Hall-MHD may require alternative formulations to the one presented here, since derived fields like the vector potential are involved.
\subsection{\label{sec:vel_units}Conservation laws with velocity units}
In closer analogy to fluid dynamics~\citep{rowley2004model}, we could have alternatively used $\bm{q} = [\bm{u},\bm{u}_A, u_s]$,
\begin{align}
    u_s^2 &= \frac{4 T}{m_i(\gamma-1)}, \quad \bm{u}_A = \frac{\bm{B}}{\sqrt{\mu_0\rho}},
\\ 
\frac{1}{2}\langle\bm{q},\bm{q}\rangle
&= \frac{1}{2}\int \left( u^2+u_A^2+u_s^2\right) d^3\bm{x}.
\end{align}
We have defined a scaled plasma sound speed, $u_s$.  
If $\rho$ is uniform 
    $\rho\langle\bm{q},\bm{q}\rangle/2 = W$.
The isothermal and time-independent density assumptions allow us to derive another quadratic model in $\bm{q}$, for which a POD-Galerkin model is readily available (the form is identical to Eq.~\ref{eq:Galerkin_model} but the coefficients have changed). Once again, assume $\bm{u}\cdot\hat{\bm{n}}= \bm{u}\times\hat{\bm{n}}=0$, and $\bm{B}\cdot\hat{\bm{n}}=0$ on the boundary, so that
\begin{align}
     \int \frac{\rho}{2}\frac{dq^2}{dt} d^3\bm{x} = \frac{\partial W}{\partial t}.
\end{align}
This is equivalent to Eq.~\eqref{eq:q2evo} in the particular case of time-independent density. Without this assumption, an extra term appears, proportional to $\int\bm{u}\cdot\nabla(u^2+u_A^2)d^3\bm{x}$. Although from dimensional analysis this term is potentially very large, this may not be the case for many laboratory devices with strong anisotropy introduced by a large external magnetic field. For instance, steady-state toroidal plasmas with large closed flux surfaces would expect $\bm{u}\cdot\nabla u_A^2$ and $\bm{u}\cdot\nabla u^2$ to be small, as the fluid velocity is primarily along field lines and gradients in both the magnetic and velocity fields are primarily across field lines. For this reason, in certain devices the use of $\bm{q} = [\bm{u},\bm{u}_A,u_s]$ could be a useful alternative to the formulation used in the main body of this work. It is possible that, in these units, the structure of the nonlinearities in the associated POD-Galerkin model may prove more amenable to analysis. Now that we have illustrated how global conservation laws manifest as structure in Galerkin models, we could compute these coefficients and evolve the subsequent model. 
However, For an explicit calculation of the model coefficients, the first and second order spatial derivatives for the MHD fields must be well-approximated in the region of experimental interest. 
In some cases, high-resolution diagnostics on experimental devices can resolve these quantities in a particular region of the plasma, but even if this data is available, computing these inner products and evaluating the nonlinear terms in the model is resource-intensive. 
Fortunately, there are hyper-reduction techniques from fluid dynamics~\cite{benner2015survey}, such as the discrete empirical interpolation method (DEIM)~\cite{chaturantabut2009discrete}, QDEIM~\cite{drmac2016new}, missing-point estimation (MPE)~\cite{astrid2008missing} and gappy POD~\cite{willcox2006unsteady,carlberg2013gnat}, which can enable efficient computations. Instead of using hyper-reduction, one can use emerging and increasingly sophisticated machine learning methods to discover Galerkin models from data. In the following section, we derive constraints on machine learning methods that guarantee the model structure we derived from global conservation laws in Sec.~\ref{sec:conservation_laws}. 
%

\section{\label{sec:sindy} Global conservation laws in machine learning model discovery}
Increasingly, machine learning techniques are allowing scientists to extract a system's governing equations of motion directly from data. Here we show how these algorithms can directly incorporate global conservation laws during the search for low-dimensional models in plasma datasets.
%
We use the sparse identification of nonlinear dynamics (SINDy) algorithm~\cite{Brunton2016pnas} to identify nonlinear reduced-order models for plasmas in the accompanying work~\cite{kaptanoglu2020physics}. 

\subsection{\label{sec:sindy_review}The constrained SINDy method}
The goal of SINDy is to identify a low-dimensional model for $\bm{a}(t)$, the vector of POD amplitudes, as a sparse linear combination of elements from a library of candidate terms $\boldsymbol{\Theta} = \begin{bmatrix}\theta_1(\bm{a}) & \theta_2(\bm{a}) & \cdots & \theta_p(\bm{a})\end{bmatrix}$:
\begin{equation}\label{Eq:SINDyExpansion}
     \frac{d}{dt}{\bm{a}} = \bm{f}(\bm{a}) \approx \boldsymbol{\Theta}(\bm{a})\boldsymbol{\Xi}.
\end{equation}
To address this combinatorically hard problem, it leverages sparse regression techniques, optimizing for the sparsest set of equations that produces an accurate fit of the data. The SINDy optimization problem is 
\begin{align}
\label{eq:constrained_sindy}
    \text{min}_{\bm{\Xi}}&||\dot{\bm{a}}-\bm{\Theta}(\bm{a})\bm{\Xi}||_2^2 + \lambda R(\bm{\Xi}), \\ \notag
    &\text{subject to} \,\,\,\,\, \bm{D}\bm{\Xi}[:] = \bm{d},
\end{align}
where $R(\bm{\Xi})$ is some regularizer, like the $L_0$ or $L_1$ norm, which promotes sparsity in $\bm{\Xi}$. Here $\bm{a},\dot{\bm{a}} \in \mathds{R}^{M\times r}$, $\bm{\Theta}(\bm{a}) \in \mathds{R}^{M\times N}$, $\bm{\Xi} \in \mathds{R}^{N\times r}$, $\bm{D} \in \mathds{R}^{N_c\times rN}$, $\bm{\Xi}[:] \in \mathds{R}^{rN}$, $\bm{d} \in \mathds{R}^{N_c}$, where $N$ is the number of candidate terms, $N_c$ is the number of constraints, and $
\bm{\Xi}[:] = 
\begin{bmatrix}
    \xi_1^{a_1} &
    \cdots &
    \xi_1^{a_r} &
    \cdots &
    \xi_N^{a_1} &
    \cdots &
    \xi_N^{a_r}
\end{bmatrix}$.

\subsection{\label{sec:sindy_coeffs}Derivation of the SINDy constraints}
In Sec.~\ref{sec:conservation_laws}, we derived model constraints from global conservation laws; our goal here is to rewrite these constraints to be compatible with the notation in Eq.~\ref{eq:constrained_sindy}. The conclusions for the global conservation of energy were: 1) no constant terms, 2) an anti-symmetry constraint on the linear part of the coefficient matrix $\bm{\Xi}$, and 3) a more complicated energy-preserving structure in the quadratic coefficients.
Consider a quadratic library in a set of $r$ modes, ordered as $\bm{\Theta}(\bm{a}) = [a_1, ..., a_r, a_1a_2,...,a_{r-1}a_r,a_1^2,...,a_r^2]$. Note that this arrangement of the polynomials in $\bm{\Theta}$ differs from Loiseau et al.~\cite{Loiseau2018jfm}, so the indexing and subscripts are also different here. First we will consider the constraint on the linear part of the Galerkin model in Eq.~\ref{eq:Galerkin_model}, or equivalently that the quadratic term $\bm{a}^T\bm{L}_0\bm{a} \approx 0$. We can rewrite this in the SINDy notation as
\begin{align}
    \label{eq:linear_constraint_deriv}
    0 = \begin{bmatrix}
    a_1 & \cdots & a_r
    \end{bmatrix}
    \begin{bmatrix}
    \xi_1^{a_1} & \cdots & \xi_r^{a_1} \\
    \vdots & \ddots & \vdots \\
    \xi_r^{a_r} & \cdots & \xi_r^{a_r} \\
    \end{bmatrix}
    \begin{bmatrix}
    a_1 \\
    \vdots \\
    a_r
    \end{bmatrix}.
\end{align}
We conclude $\xi_i^{a_j} = -\xi_j^{a_i}$ for $i,j \in \{1,...,r\}$ and identify $\xi_i^{a_j}$ by accessing the $(i-1)r+j$ index in $\bm{\Xi}[:]$. Note we are only accessing the first $r^2$ elements of $\bm{\Xi}[:]$. For models limited to linear and quadratic polynomials, $N = (r^2+3r)/2$ and the number of constraints from anti-symmetry of the linear coefficients is $N_L = (r^2+r)/2$. Thus there are now only $rN - N_L = r(r^2+2r-1)/2$ free parameters. Since the constrained SINDy algorithm solves linear equality constraints of the form $\bm{D}\bm{\Xi}[:] = \bm{d}$, we can write this out explicitly for $r=3$,
\setcounter{MaxMatrixCols}{30}
\setlength{\arraycolsep}{3.0pt}
\medmuskip = 4.0mu 
\begin{align}
    \begin{bmatrix}
    \bm{1} & 0 & 0 & 0 & 0 & 0 & 0 & 0 & 0 & 0 & \cdots \\
    0 & 0 & 0 & 0 & \bm{1} & 0 & 0 & 0 & 0 & 0 & \cdots \\
    0 & 0 & 0 & 0 & 0 & 0 & 0 & 0 & \bm{1} & 0 & \cdots \\
    0 & \bm{1} & 0 & \bm{1} & 0 & 0 & 0 & 0 & 0 & 0 & \cdots \\
    0 & 0 & \bm{1} & 0 & 0 & 0 & \bm{1} & 0 & 0 & 0 & \cdots \\
    0 & 0 & 0 & 0 & 0 & \bm{1} & 0 & \bm{1} & 0 & 0 & \cdots \\
    \end{bmatrix}\begin{bmatrix}
    \xi_1^{a_1} \\
    \xi_1^{a_2} \\
    \xi_1^{a_3} \\
    \xi_2^{a_1} \\
    \vdots \\
\end{bmatrix}
= \begin{bmatrix}
    0 \\ 
    0 \\
    0 \\
    0 \\
    0 \\
    0 \\
\end{bmatrix}.
\end{align}
\medmuskip = 0mu 
The boundary conditions $\bm{u}\cdot\hat{\bm{n}} = 0$, $\bm{J}\cdot\hat{\bm{n}} = 0$, $\bm{B}\cdot\hat{\bm{n}} = 0$ guaranteed that the quadratic nonlinearities were energy-preserving, and thus that cubic terms in Eq.~\ref{eq:galtier_deriv} vanish. In this case, regardless of the steady-state assumption, we have that
\begin{align}
    \sum_{i,j,k=0}^rQ_{ijk}^0a_ia_ja_k \approx 0.
\end{align}
This constraint is significantly more involved to reformat. Written in SINDy notation, this is equivalent to
\begin{align}
\label{eq:quad_constraint_formulation}
    0 =\begin{bmatrix}
        a_1 & \cdots & a_r
    \end{bmatrix}
    \begin{bmatrix}
        \xi_{r+1}^{a_1} & \xi_{r+2}^{a_1} & \cdots & \xi_{N}^{a_1} \\
        \vdots & \vdots & \vdots & \vdots \\
        \xi_{r+1}^{a_r} & \xi_{r+2}^{a_r} & \cdots & \xi_{N}^{a_r}
    \end{bmatrix}
    \begin{bmatrix}
        a_1a_2 \\
        \vdots \\
        a_{r-1}a_r \\
        a_1^2 \\
        \vdots \\
        a_r^2
    \end{bmatrix}.
\end{align}
Expand this all out and group the like terms, i.e. terms which look like $a_i^3$, $a_ia_j^2$ or $a_ia_ja_k$, $i,j,k \in \{1,...,r\}$, $i\neq j \neq k$. All of the like terms can be straightforwardly shown to be linearly independent, so we can consider three constraints separately for the three types of terms. The number of each of these respective terms is ${r \choose 1} = r$, 2${r \choose 2} = r(r-1)$, and ${r \choose 3} = r(r-1)(r-2)/6$, for a total of $r(r+1)(r+2)/6 = N_Q$ constraints. With both constraints, we have $rN-N_L-N_Q = r(r-1)(2r+5)/6$ free parameters, and $N_c = N_L + N_Q$ constraints. 
%
Further considering the quadratic case, we find that coefficients which adorn $a_i^3$ must vanish, $
\xi_{N-r+i}^{a_{i}} = 0$. 
Now define
\begin{equation}
    \tilde{\xi}_{ijk} = \xi_{r+\frac{j}{2}(2r-j-3)+k-1}^{a_i}.
\end{equation}
The second type of constraint, with $i\neq j$, produces 
\begin{equation}
\xi_{N-r+j}^{a_i} =\begin{cases}
 \tilde{\xi}_{jij} & i < j \\
\tilde{\xi}_{jji} & i > j,
\end{cases}
\end{equation}
while the third type of constraint produces 
\begin{equation}
    \label{eq:ai_aj_ak}
\tilde{\xi}_{ijk} + \tilde{\xi}_{jik} + \tilde{\xi}_{kij} = 0.
\end{equation}
This relation is equivalent to the energy-preserving conditions in Schlegel et al.~\cite{Schlegel2015jfm}, but the indexing is not straightforward, even after fully expanding Eq.~\ref{eq:quad_constraint_formulation}. 
This equation is an arbitrary $r$ generalization to the $r=3$ constraint used in Loiseau et al.~\cite{Loiseau2018jfm}. 
For the specific case where the plasma system is Hamiltonian (for instance in ideal~\cite{morrison1980noncanonical}, Hall~\cite{yoshida2013canonical}, and extended~\citep{abdelhamid2015hamiltonian} MHD without dissipation) and the measurements are assumed to be sufficient to represent the Hamiltonian, one could alternatively use formulations of SINDy to directly discover the Hamiltonian~\cite{chu2020discovering} and subsequently derive the equations of motion.
Lastly, cross-helicity constraints can be straightforwardly implemented from these results. If the constraint on the quadratic terms in $\bm{\Xi}$ is written $D_{jk}\Xi_k = 0$, then the quadratic cross-helicity constraint can be written $D_{jk}A_{kl}\Xi_l = 0$. 
\section{Conclusion}\label{sec:conclusion}
A hierarchy of models with varying fidelity is essential for understanding and controlling plasmas. The present work, along with the accompanying work~\cite{kaptanoglu2020physics}, provides a principled lower level on this hierarchy $-$ low-dimensional and interpretable plasma models which can be used for physical discovery, forecasting, stability analysis, and real-time control.
We have illustrated how Galerkin plasma models retain the global conservation laws of MHD, and how machine learning methods like SINDy can use these constraints directly in an optimization procedure for discovering such models from data. 
This principled enforcement of global conservation laws is critical for the success of future low-dimensional plasma models. 

\section{Acknowledgements\label{sec:acknowledgements}}
The authors would like to extend their gratitude to Dr. Uri Shumlak for his input on this work. This work and the companion work were supported by the Army Research Office ({ARO W}911{NF}-19-1-0045) and the Air Force Office of Scientific Research (AFOSR {FA}9550-18-1-0200). Simulations were supported by the U.S. Department of Energy under award numbers {DE-SC}0016256 and DE-AR0001098 and used resources of the National Energy Research Scientific Computing Center, supported by the Office of Science of the U.S. Department of Energy under Contract No. DE-AC02–05CH11231.


\newpage
 
 \bibliography{Galerkin}

\begin{thebibliography}{51}%
\makeatletter
\providecommand \@ifxundefined [1]{%
 \@ifx{#1\undefined}
}%
\providecommand \@ifnum [1]{%
 \ifnum #1\expandafter \@firstoftwo
 \else \expandafter \@secondoftwo
 \fi
}%
\providecommand \@ifx [1]{%
 \ifx #1\expandafter \@firstoftwo
 \else \expandafter \@secondoftwo
 \fi
}%
\providecommand \natexlab [1]{#1}%
\providecommand \enquote  [1]{``#1''}%
\providecommand \bibnamefont  [1]{#1}%
\providecommand \bibfnamefont [1]{#1}%
\providecommand \citenamefont [1]{#1}%
\providecommand \href@noop [0]{\@secondoftwo}%
\providecommand \href [0]{\begingroup \@sanitize@url \@href}%
\providecommand \@href[1]{\@@startlink{#1}\@@href}%
\providecommand \@@href[1]{\endgroup#1\@@endlink}%
\providecommand \@sanitize@url [0]{\catcode `\\12\catcode `\$12\catcode
  `\&12\catcode `\#12\catcode `\^12\catcode `\_12\catcode `\%12\relax}%
\providecommand \@@startlink[1]{}%
\providecommand \@@endlink[0]{}%
\providecommand \url  [0]{\begingroup\@sanitize@url \@url }%
\providecommand \@url [1]{\endgroup\@href {#1}{\urlprefix }}%
\providecommand \urlprefix  [0]{URL }%
\providecommand \Eprint [0]{\href }%
\providecommand \doibase [0]{https://doi.org/}%
\providecommand \selectlanguage [0]{\@gobble}%
\providecommand \bibinfo  [0]{\@secondoftwo}%
\providecommand \bibfield  [0]{\@secondoftwo}%
\providecommand \translation [1]{[#1]}%
\providecommand \BibitemOpen [0]{}%
\providecommand \bibitemStop [0]{}%
\providecommand \bibitemNoStop [0]{.\EOS\space}%
\providecommand \EOS [0]{\spacefactor3000\relax}%
\providecommand \BibitemShut  [1]{\csname bibitem#1\endcsname}%
\let\auto@bib@innerbib\@empty
\bibitem [{\citenamefont {Candy}\ and\ \citenamefont
  {Waltz}(2003)}]{Candy2003}%
  \BibitemOpen
  \bibfield  {author} {\bibinfo {author} {\bibfnamefont {J.}~\bibnamefont
  {Candy}}\ and\ \bibinfo {author} {\bibfnamefont {R.~E.}\ \bibnamefont
  {Waltz}},\ }\bibfield  {title} {\bibinfo {title} {Anomalous transport scaling
  in the {DIII-D} tokamak matched by supercomputer simulation},\ }\href
  {https://doi.org/10.1103/PhysRevLett.91.045001} {\bibfield  {journal}
  {\bibinfo  {journal} {Phys. Rev. Lett.}\ }\textbf {\bibinfo {volume} {91}},\
  \bibinfo {pages} {045001} (\bibinfo {year} {2003})}\BibitemShut {NoStop}%
\bibitem [{\citenamefont {Ohia}\ \emph {et~al.}(2012)\citenamefont {Ohia},
  \citenamefont {Egedal}, \citenamefont {Lukin}, \citenamefont {Daughton},\
  and\ \citenamefont {Le}}]{Ohia2012}%
  \BibitemOpen
  \bibfield  {author} {\bibinfo {author} {\bibfnamefont {O.}~\bibnamefont
  {Ohia}}, \bibinfo {author} {\bibfnamefont {J.}~\bibnamefont {Egedal}},
  \bibinfo {author} {\bibfnamefont {V.~S.}\ \bibnamefont {Lukin}}, \bibinfo
  {author} {\bibfnamefont {W.}~\bibnamefont {Daughton}},\ and\ \bibinfo
  {author} {\bibfnamefont {A.}~\bibnamefont {Le}},\ }\bibfield  {title}
  {\bibinfo {title} {Demonstration of anisotropic fluid closure capturing the
  kinetic structure of magnetic reconnection},\ }\href
  {https://doi.org/10.1103/PhysRevLett.109.115004} {\bibfield  {journal}
  {\bibinfo  {journal} {Phys. Rev. Lett.}\ }\textbf {\bibinfo {volume} {109}},\
  \bibinfo {pages} {115004} (\bibinfo {year} {2012})}\BibitemShut {NoStop}%
\bibitem [{\citenamefont {Gro\ifmmode~\check{s}\else \v{s}\fi{}elj}\ \emph
  {et~al.}(2018)\citenamefont {Gro\ifmmode~\check{s}\else \v{s}\fi{}elj},
  \citenamefont {Mallet}, \citenamefont {Loureiro},\ and\ \citenamefont
  {Jenko}}]{Groselj2018}%
  \BibitemOpen
  \bibfield  {author} {\bibinfo {author} {\bibfnamefont {D.}~\bibnamefont
  {Gro\ifmmode~\check{s}\else \v{s}\fi{}elj}}, \bibinfo {author} {\bibfnamefont
  {A.}~\bibnamefont {Mallet}}, \bibinfo {author} {\bibfnamefont {N.~F.}\
  \bibnamefont {Loureiro}},\ and\ \bibinfo {author} {\bibfnamefont
  {F.}~\bibnamefont {Jenko}},\ }\bibfield  {title} {\bibinfo {title} {Fully
  kinetic simulation of {3D} kinetic {A}lfv\'en turbulence},\ }\href
  {https://doi.org/10.1103/PhysRevLett.120.105101} {\bibfield  {journal}
  {\bibinfo  {journal} {Phys. Rev. Lett.}\ }\textbf {\bibinfo {volume} {120}},\
  \bibinfo {pages} {105101} (\bibinfo {year} {2018})}\BibitemShut {NoStop}%
\bibitem [{\citenamefont {Taira}\ \emph {et~al.}(2017)\citenamefont {Taira},
  \citenamefont {Brunton}, \citenamefont {Dawson}, \citenamefont {Rowley},
  \citenamefont {Colonius}, \citenamefont {McKeon}, \citenamefont {Schmidt},
  \citenamefont {Gordeyev}, \citenamefont {Theofilis},\ and\ \citenamefont
  {Ukeiley}}]{Taira2017aiaa}%
  \BibitemOpen
  \bibfield  {author} {\bibinfo {author} {\bibfnamefont {K.}~\bibnamefont
  {Taira}}, \bibinfo {author} {\bibfnamefont {S.~L.}\ \bibnamefont {Brunton}},
  \bibinfo {author} {\bibfnamefont {S.}~\bibnamefont {Dawson}}, \bibinfo
  {author} {\bibfnamefont {C.~W.}\ \bibnamefont {Rowley}}, \bibinfo {author}
  {\bibfnamefont {T.}~\bibnamefont {Colonius}}, \bibinfo {author}
  {\bibfnamefont {B.~J.}\ \bibnamefont {McKeon}}, \bibinfo {author}
  {\bibfnamefont {O.~T.}\ \bibnamefont {Schmidt}}, \bibinfo {author}
  {\bibfnamefont {S.}~\bibnamefont {Gordeyev}}, \bibinfo {author}
  {\bibfnamefont {V.}~\bibnamefont {Theofilis}},\ and\ \bibinfo {author}
  {\bibfnamefont {L.~S.}\ \bibnamefont {Ukeiley}},\ }\bibfield  {title}
  {\bibinfo {title} {Modal analysis of fluid flows: An overview},\ }\href@noop
  {} {\bibfield  {journal} {\bibinfo  {journal} {AIAA Journal}\ }\textbf
  {\bibinfo {volume} {55}},\ \bibinfo {pages} {4013} (\bibinfo {year}
  {2017})}\BibitemShut {NoStop}%
\bibitem [{\citenamefont {Furno}(2001)}]{furno2001fast}%
  \BibitemOpen
  \bibfield  {author} {\bibinfo {author} {\bibfnamefont {I.}~\bibnamefont
  {Furno}},\ }\href@noop {} {\emph {\bibinfo {title} {Fast transient transport
  phenomena measured by soft x-ray emission in {TCV} tokamak plasmas}}},\
  \bibinfo {type} {Tech. Rep.}\ (\bibinfo {year} {2001})\BibitemShut {NoStop}%
\bibitem [{\citenamefont {Jim{\'e}nez-G{\'o}mez}\ \emph
  {et~al.}(2007)\citenamefont {Jim{\'e}nez-G{\'o}mez}, \citenamefont
  {Ascas{\'\i}bar}, \citenamefont {Estrada}, \citenamefont
  {Garc{\'\i}a-Cort{\'e}s}, \citenamefont {Van~Milligen}, \citenamefont
  {L{\'o}pez-Fraguas}, \citenamefont {Pastor},\ and\ \citenamefont
  {L{\'o}pez-Bruna}}]{jimenez2007analysis}%
  \BibitemOpen
  \bibfield  {author} {\bibinfo {author} {\bibfnamefont {R.}~\bibnamefont
  {Jim{\'e}nez-G{\'o}mez}}, \bibinfo {author} {\bibfnamefont {E.}~\bibnamefont
  {Ascas{\'\i}bar}}, \bibinfo {author} {\bibfnamefont {T.}~\bibnamefont
  {Estrada}}, \bibinfo {author} {\bibfnamefont {I.}~\bibnamefont
  {Garc{\'\i}a-Cort{\'e}s}}, \bibinfo {author} {\bibfnamefont {B.}~\bibnamefont
  {Van~Milligen}}, \bibinfo {author} {\bibfnamefont {A.}~\bibnamefont
  {L{\'o}pez-Fraguas}}, \bibinfo {author} {\bibfnamefont {I.}~\bibnamefont
  {Pastor}},\ and\ \bibinfo {author} {\bibfnamefont {D.}~\bibnamefont
  {L{\'o}pez-Bruna}},\ }\bibfield  {title} {\bibinfo {title} {Analysis of
  magnetohydrodynamic instabilities in {TJ-II} plasmas},\ }\href@noop {}
  {\bibfield  {journal} {\bibinfo  {journal} {Fusion science and technology}\
  }\textbf {\bibinfo {volume} {51}},\ \bibinfo {pages} {20} (\bibinfo {year}
  {2007})}\BibitemShut {NoStop}%
\bibitem [{\citenamefont {van Milligen}\ \emph {et~al.}(2014)\citenamefont {van
  Milligen}, \citenamefont {S{\'{a}}nchez}, \citenamefont {Alonso},
  \citenamefont {Pedrosa}, \citenamefont {Hidalgo}, \citenamefont
  {de~Aguilera},\ and\ \citenamefont {Fraguas}}]{vanMilligen14}%
  \BibitemOpen
  \bibfield  {author} {\bibinfo {author} {\bibfnamefont {B.~P.}\ \bibnamefont
  {van Milligen}}, \bibinfo {author} {\bibfnamefont {E.}~\bibnamefont
  {S{\'{a}}nchez}}, \bibinfo {author} {\bibfnamefont {A.}~\bibnamefont
  {Alonso}}, \bibinfo {author} {\bibfnamefont {M.~A.}\ \bibnamefont {Pedrosa}},
  \bibinfo {author} {\bibfnamefont {C.}~\bibnamefont {Hidalgo}}, \bibinfo
  {author} {\bibfnamefont {A.~M.}\ \bibnamefont {de~Aguilera}},\ and\ \bibinfo
  {author} {\bibfnamefont {A.~L.}\ \bibnamefont {Fraguas}},\ }\bibfield
  {title} {\bibinfo {title} {The use of the biorthogonal decomposition for the
  identification of zonal flows at {TJ}-{II}},\ }\href@noop {} {\bibfield
  {journal} {\bibinfo  {journal} {Plasma Physics and Controlled Fusion}\
  }\textbf {\bibinfo {volume} {57}},\ \bibinfo {pages} {025005} (\bibinfo
  {year} {2014})}\BibitemShut {NoStop}%
\bibitem [{\citenamefont {Pandya}(2016)}]{pandya}%
  \BibitemOpen
  \bibfield  {author} {\bibinfo {author} {\bibfnamefont {M.}~\bibnamefont
  {Pandya}},\ }\emph {\bibinfo {title} {Low edge safety factor disruptions in
  the Compact Toroidal Hybrid: Operation in the low-q regime, passive
  disruption avoidance and the nature of {MHD} precursors}},\ \href@noop {}
  {Ph.D. thesis},\ \bibinfo  {school} {{A}uburn {U}niversity} (\bibinfo {year}
  {2016})\BibitemShut {NoStop}%
\bibitem [{\citenamefont {Victor}\ \emph {et~al.}(2015)\citenamefont {Victor},
  \citenamefont {Akcay}, \citenamefont {Hansen}, \citenamefont {Jarboe},
  \citenamefont {Nelson},\ and\ \citenamefont
  {Morgan}}]{victor2015development}%
  \BibitemOpen
  \bibfield  {author} {\bibinfo {author} {\bibfnamefont {B.}~\bibnamefont
  {Victor}}, \bibinfo {author} {\bibfnamefont {C.}~\bibnamefont {Akcay}},
  \bibinfo {author} {\bibfnamefont {C.}~\bibnamefont {Hansen}}, \bibinfo
  {author} {\bibfnamefont {T.}~\bibnamefont {Jarboe}}, \bibinfo {author}
  {\bibfnamefont {B.}~\bibnamefont {Nelson}},\ and\ \bibinfo {author}
  {\bibfnamefont {K.}~\bibnamefont {Morgan}},\ }\bibfield  {title} {\bibinfo
  {title} {Development of validation metrics using biorthogonal decomposition
  for the comparison of magnetic field measurements},\ }\href@noop {}
  {\bibfield  {journal} {\bibinfo  {journal} {Plasma Physics and Controlled
  Fusion}\ }\textbf {\bibinfo {volume} {57}},\ \bibinfo {pages} {045010}
  (\bibinfo {year} {2015})}\BibitemShut {NoStop}%
\bibitem [{\citenamefont {Strait}\ \emph {et~al.}(2016)\citenamefont {Strait},
  \citenamefont {King}, \citenamefont {Hanson},\ and\ \citenamefont
  {Logan}}]{strait2016spatial}%
  \BibitemOpen
  \bibfield  {author} {\bibinfo {author} {\bibfnamefont {E.}~\bibnamefont
  {Strait}}, \bibinfo {author} {\bibfnamefont {J.}~\bibnamefont {King}},
  \bibinfo {author} {\bibfnamefont {J.}~\bibnamefont {Hanson}},\ and\ \bibinfo
  {author} {\bibfnamefont {N.}~\bibnamefont {Logan}},\ }\bibfield  {title}
  {\bibinfo {title} {Spatial and temporal analysis of {DIII-D 3D} magnetic
  diagnostic data},\ }\href@noop {} {\bibfield  {journal} {\bibinfo  {journal}
  {Review of Scientific Instruments}\ }\textbf {\bibinfo {volume} {87}},\
  \bibinfo {pages} {11D423} (\bibinfo {year} {2016})}\BibitemShut {NoStop}%
\bibitem [{\citenamefont {Byrne}(2017)}]{byrne2017study}%
  \BibitemOpen
  \bibfield  {author} {\bibinfo {author} {\bibfnamefont {P.~J.}\ \bibnamefont
  {Byrne}},\ }\emph {\bibinfo {title} {Study of External Kink Modes in Shaped
  HBT-EP Plasmas}},\ \href@noop {} {Ph.D. thesis},\ \bibinfo  {school}
  {Columbia University} (\bibinfo {year} {2017})\BibitemShut {NoStop}%
\bibitem [{\citenamefont {Gu}\ \emph {et~al.}(2019)\citenamefont {Gu},
  \citenamefont {Wan}, \citenamefont {Sun}, \citenamefont {Chu}, \citenamefont
  {Liu}, \citenamefont {Shi}, \citenamefont {Wang}, \citenamefont {Jia},\ and\
  \citenamefont {He}}]{gu2019new}%
  \BibitemOpen
  \bibfield  {author} {\bibinfo {author} {\bibfnamefont {S.}~\bibnamefont
  {Gu}}, \bibinfo {author} {\bibfnamefont {B.}~\bibnamefont {Wan}}, \bibinfo
  {author} {\bibfnamefont {Y.}~\bibnamefont {Sun}}, \bibinfo {author}
  {\bibfnamefont {N.}~\bibnamefont {Chu}}, \bibinfo {author} {\bibfnamefont
  {Y.}~\bibnamefont {Liu}}, \bibinfo {author} {\bibfnamefont {T.}~\bibnamefont
  {Shi}}, \bibinfo {author} {\bibfnamefont {H.}~\bibnamefont {Wang}}, \bibinfo
  {author} {\bibfnamefont {M.}~\bibnamefont {Jia}},\ and\ \bibinfo {author}
  {\bibfnamefont {K.}~\bibnamefont {He}},\ }\bibfield  {title} {\bibinfo
  {title} {A new criterion for controlling edge localized modes based on a
  multi-mode plasma response},\ }\href@noop {} {\bibfield  {journal} {\bibinfo
  {journal} {Nuclear Fusion}\ }\textbf {\bibinfo {volume} {59}},\ \bibinfo
  {pages} {126042} (\bibinfo {year} {2019})}\BibitemShut {NoStop}%
\bibitem [{\citenamefont {Kaptanoglu}\ \emph
  {et~al.}(2020{\natexlab{a}})\citenamefont {Kaptanoglu}, \citenamefont
  {Morgan}, \citenamefont {Hansen},\ and\ \citenamefont
  {Brunton}}]{kaptanoglu2020}%
  \BibitemOpen
  \bibfield  {author} {\bibinfo {author} {\bibfnamefont {A.~A.}\ \bibnamefont
  {Kaptanoglu}}, \bibinfo {author} {\bibfnamefont {K.~D.}\ \bibnamefont
  {Morgan}}, \bibinfo {author} {\bibfnamefont {C.~J.}\ \bibnamefont {Hansen}},\
  and\ \bibinfo {author} {\bibfnamefont {S.~L.}\ \bibnamefont {Brunton}},\
  }\bibfield  {title} {\bibinfo {title} {Characterizing magnetized plasmas with
  dynamic mode decomposition},\ }\href@noop {} {\bibfield  {journal} {\bibinfo
  {journal} {Physics of Plasmas}\ }\textbf {\bibinfo {volume} {27}},\ \bibinfo
  {pages} {032108} (\bibinfo {year} {2020}{\natexlab{a}})}\BibitemShut
  {NoStop}%
\bibitem [{\citenamefont {Benner}\ \emph {et~al.}(2005)\citenamefont {Benner},
  \citenamefont {Mehrmann},\ and\ \citenamefont
  {Sorensen}}]{benner2005dimension}%
  \BibitemOpen
  \bibfield  {author} {\bibinfo {author} {\bibfnamefont {P.}~\bibnamefont
  {Benner}}, \bibinfo {author} {\bibfnamefont {V.}~\bibnamefont {Mehrmann}},\
  and\ \bibinfo {author} {\bibfnamefont {D.~C.}\ \bibnamefont {Sorensen}},\
  }\href@noop {} {\emph {\bibinfo {title} {Dimension reduction of large-scale
  systems}}},\ Vol.~\bibinfo {volume} {45}\ (\bibinfo  {publisher} {Springer},\
  \bibinfo {year} {2005})\BibitemShut {NoStop}%
\bibitem [{\citenamefont {Benner}\ \emph {et~al.}(2017)\citenamefont {Benner},
  \citenamefont {Ohlberger}, \citenamefont {Cohen},\ and\ \citenamefont
  {Willcox}}]{benner2017model}%
  \BibitemOpen
  \bibfield  {author} {\bibinfo {author} {\bibfnamefont {P.}~\bibnamefont
  {Benner}}, \bibinfo {author} {\bibfnamefont {M.}~\bibnamefont {Ohlberger}},
  \bibinfo {author} {\bibfnamefont {A.}~\bibnamefont {Cohen}},\ and\ \bibinfo
  {author} {\bibfnamefont {K.}~\bibnamefont {Willcox}},\ }\href@noop {} {\emph
  {\bibinfo {title} {Model reduction and approximation: theory and
  algorithms}}}\ (\bibinfo  {publisher} {SIAM},\ \bibinfo {year}
  {2017})\BibitemShut {NoStop}%
\bibitem [{\citenamefont {Kaptanoglu}\ \emph
  {et~al.}(2020{\natexlab{b}})\citenamefont {Kaptanoglu}, \citenamefont
  {Morgan}, \citenamefont {Hansen},\ and\ \citenamefont
  {Brunton}}]{kaptanoglu2020physics}%
  \BibitemOpen
  \bibfield  {author} {\bibinfo {author} {\bibfnamefont {A.~A.}\ \bibnamefont
  {Kaptanoglu}}, \bibinfo {author} {\bibfnamefont {K.~D.}\ \bibnamefont
  {Morgan}}, \bibinfo {author} {\bibfnamefont {C.~J.}\ \bibnamefont {Hansen}},\
  and\ \bibinfo {author} {\bibfnamefont {S.~L.}\ \bibnamefont {Brunton}},\
  }\bibfield  {title} {\bibinfo {title} {Physics-constrained, low-dimensional
  models for mhd: First-principles and data-driven approaches},\ }\href@noop {}
  {\bibfield  {journal} {\bibinfo  {journal} {arXiv preprint arXiv:2004.10389}\
  } (\bibinfo {year} {2020}{\natexlab{b}})}\BibitemShut {NoStop}%
\bibitem [{\citenamefont {Rowley}\ \emph {et~al.}(2004)\citenamefont {Rowley},
  \citenamefont {Colonius},\ and\ \citenamefont {Murray}}]{rowley2004model}%
  \BibitemOpen
  \bibfield  {author} {\bibinfo {author} {\bibfnamefont {C.~W.}\ \bibnamefont
  {Rowley}}, \bibinfo {author} {\bibfnamefont {T.}~\bibnamefont {Colonius}},\
  and\ \bibinfo {author} {\bibfnamefont {R.~M.}\ \bibnamefont {Murray}},\
  }\bibfield  {title} {\bibinfo {title} {Model reduction for compressible flows
  using {POD} and {G}alerkin projection},\ }\href@noop {} {\bibfield  {journal}
  {\bibinfo  {journal} {Physica D: Nonlinear Phenomena}\ }\textbf {\bibinfo
  {volume} {189}},\ \bibinfo {pages} {115} (\bibinfo {year}
  {2004})}\BibitemShut {NoStop}%
\bibitem [{\citenamefont {Noack}\ \emph {et~al.}(2011)\citenamefont {Noack},
  \citenamefont {Schlegel}, \citenamefont {Morzynski},\ and\ \citenamefont
  {Tadmor}}]{noack2011galerkin}%
  \BibitemOpen
  \bibfield  {author} {\bibinfo {author} {\bibfnamefont {B.~R.}\ \bibnamefont
  {Noack}}, \bibinfo {author} {\bibfnamefont {M.}~\bibnamefont {Schlegel}},
  \bibinfo {author} {\bibfnamefont {M.}~\bibnamefont {Morzynski}},\ and\
  \bibinfo {author} {\bibfnamefont {G.}~\bibnamefont {Tadmor}},\ }\href@noop {}
  {\emph {\bibinfo {title} {Galerkin method for nonlinear dynamics}}}\
  (\bibinfo  {publisher} {Springer},\ \bibinfo {year} {2011})\BibitemShut
  {NoStop}%
\bibitem [{\citenamefont {Balajewicz}\ \emph {et~al.}(2013)\citenamefont
  {Balajewicz}, \citenamefont {Dowell},\ and\ \citenamefont
  {Noack}}]{balajewicz2013low}%
  \BibitemOpen
  \bibfield  {author} {\bibinfo {author} {\bibfnamefont {M.~J.}\ \bibnamefont
  {Balajewicz}}, \bibinfo {author} {\bibfnamefont {E.~H.}\ \bibnamefont
  {Dowell}},\ and\ \bibinfo {author} {\bibfnamefont {B.~R.}\ \bibnamefont
  {Noack}},\ }\bibfield  {title} {\bibinfo {title} {Low-dimensional modelling
  of high-{R}eynolds-number shear flows incorporating constraints from the
  {N}avier--{S}tokes equation},\ }\href@noop {} {\bibfield  {journal} {\bibinfo
   {journal} {Journal of Fluid Mechanics}\ }\textbf {\bibinfo {volume} {729}},\
  \bibinfo {pages} {285} (\bibinfo {year} {2013})}\BibitemShut {NoStop}%
\bibitem [{\citenamefont {Schlegel}\ and\ \citenamefont
  {Noack}(2015)}]{Schlegel2015jfm}%
  \BibitemOpen
  \bibfield  {author} {\bibinfo {author} {\bibfnamefont {M.}~\bibnamefont
  {Schlegel}}\ and\ \bibinfo {author} {\bibfnamefont {B.~R.}\ \bibnamefont
  {Noack}},\ }\bibfield  {title} {\bibinfo {title} {On long-term boundedness of
  {G}alerkin models},\ }\href@noop {} {\bibfield  {journal} {\bibinfo
  {journal} {Journal of Fluid Mechanics}\ }\textbf {\bibinfo {volume} {765}},\
  \bibinfo {pages} {325} (\bibinfo {year} {2015})}\BibitemShut {NoStop}%
\bibitem [{\citenamefont {Carlberg}\ \emph {et~al.}(2017)\citenamefont
  {Carlberg}, \citenamefont {Barone},\ and\ \citenamefont
  {Antil}}]{Carlberg2017jcp}%
  \BibitemOpen
  \bibfield  {author} {\bibinfo {author} {\bibfnamefont {K.}~\bibnamefont
  {Carlberg}}, \bibinfo {author} {\bibfnamefont {M.}~\bibnamefont {Barone}},\
  and\ \bibinfo {author} {\bibfnamefont {H.}~\bibnamefont {Antil}},\ }\bibfield
   {title} {\bibinfo {title} {Galerkin v. least-squares {P}etrov--{G}alerkin
  projection in nonlinear model reduction},\ }\href@noop {} {\bibfield
  {journal} {\bibinfo  {journal} {Journal of Computational Physics}\ }\textbf
  {\bibinfo {volume} {330}},\ \bibinfo {pages} {693} (\bibinfo {year}
  {2017})}\BibitemShut {NoStop}%
\bibitem [{\citenamefont {Dudok~de Wit}\ \emph {et~al.}(1994)\citenamefont
  {Dudok~de Wit}, \citenamefont {Pecquet}, \citenamefont {Vallet},\ and\
  \citenamefont {Lima}}]{dudok1994biorthogonal}%
  \BibitemOpen
  \bibfield  {author} {\bibinfo {author} {\bibfnamefont {T.}~\bibnamefont
  {Dudok~de Wit}}, \bibinfo {author} {\bibfnamefont {A.-L.}\ \bibnamefont
  {Pecquet}}, \bibinfo {author} {\bibfnamefont {J.-C.}\ \bibnamefont
  {Vallet}},\ and\ \bibinfo {author} {\bibfnamefont {R.}~\bibnamefont {Lima}},\
  }\bibfield  {title} {\bibinfo {title} {The biorthogonal decomposition as a
  tool for investigating fluctuations in plasmas},\ }\href@noop {} {\bibfield
  {journal} {\bibinfo  {journal} {Physics of Plasmas}\ }\textbf {\bibinfo
  {volume} {1}},\ \bibinfo {pages} {3288} (\bibinfo {year} {1994})}\BibitemShut
  {NoStop}%
\bibitem [{\citenamefont {Levesque}\ \emph {et~al.}(2013)\citenamefont
  {Levesque}, \citenamefont {Rath}, \citenamefont {Shiraki}, \citenamefont
  {Angelini}, \citenamefont {Bialek}, \citenamefont {Byrne}, \citenamefont
  {DeBono}, \citenamefont {Hughes}, \citenamefont {Mauel}, \citenamefont
  {Navratil} \emph {et~al.}}]{levesque2013multimode}%
  \BibitemOpen
  \bibfield  {author} {\bibinfo {author} {\bibfnamefont {J.}~\bibnamefont
  {Levesque}}, \bibinfo {author} {\bibfnamefont {N.}~\bibnamefont {Rath}},
  \bibinfo {author} {\bibfnamefont {D.}~\bibnamefont {Shiraki}}, \bibinfo
  {author} {\bibfnamefont {S.}~\bibnamefont {Angelini}}, \bibinfo {author}
  {\bibfnamefont {J.}~\bibnamefont {Bialek}}, \bibinfo {author} {\bibfnamefont
  {P.}~\bibnamefont {Byrne}}, \bibinfo {author} {\bibfnamefont
  {B.}~\bibnamefont {DeBono}}, \bibinfo {author} {\bibfnamefont
  {P.}~\bibnamefont {Hughes}}, \bibinfo {author} {\bibfnamefont
  {M.}~\bibnamefont {Mauel}}, \bibinfo {author} {\bibfnamefont
  {G.}~\bibnamefont {Navratil}}, \emph {et~al.},\ }\bibfield  {title} {\bibinfo
  {title} {Multimode observations and 3{D} magnetic control of the boundary of
  a tokamak plasma},\ }\href@noop {} {\bibfield  {journal} {\bibinfo  {journal}
  {Nuclear Fusion}\ }\textbf {\bibinfo {volume} {53}},\ \bibinfo {pages}
  {073037} (\bibinfo {year} {2013})}\BibitemShut {NoStop}%
\bibitem [{\citenamefont {Galperti}\ \emph {et~al.}(2014)\citenamefont
  {Galperti}, \citenamefont {Marchetto}, \citenamefont {Alessi}, \citenamefont
  {Minelli}, \citenamefont {Mosconi}, \citenamefont {Belli}, \citenamefont
  {Boncagni}, \citenamefont {Botrugno}, \citenamefont {Buratti}, \citenamefont
  {Esposito} \emph {et~al.}}]{galperti2014development}%
  \BibitemOpen
  \bibfield  {author} {\bibinfo {author} {\bibfnamefont {C.}~\bibnamefont
  {Galperti}}, \bibinfo {author} {\bibfnamefont {C.}~\bibnamefont {Marchetto}},
  \bibinfo {author} {\bibfnamefont {E.}~\bibnamefont {Alessi}}, \bibinfo
  {author} {\bibfnamefont {D.}~\bibnamefont {Minelli}}, \bibinfo {author}
  {\bibfnamefont {M.}~\bibnamefont {Mosconi}}, \bibinfo {author} {\bibfnamefont
  {F.}~\bibnamefont {Belli}}, \bibinfo {author} {\bibfnamefont
  {L.}~\bibnamefont {Boncagni}}, \bibinfo {author} {\bibfnamefont
  {A.}~\bibnamefont {Botrugno}}, \bibinfo {author} {\bibfnamefont
  {P.}~\bibnamefont {Buratti}}, \bibinfo {author} {\bibfnamefont
  {B.}~\bibnamefont {Esposito}}, \emph {et~al.},\ }\bibfield  {title} {\bibinfo
  {title} {Development of real-time {MHD} markers based on biorthogonal
  decomposition of signals from {M}irnov coils},\ }\href@noop {} {\bibfield
  {journal} {\bibinfo  {journal} {Plasma Physics and Controlled Fusion}\
  }\textbf {\bibinfo {volume} {56}},\ \bibinfo {pages} {114012} (\bibinfo
  {year} {2014})}\BibitemShut {NoStop}%
\bibitem [{\citenamefont {Van~Milligen}\ \emph {et~al.}(2014)\citenamefont
  {Van~Milligen}, \citenamefont {S{\'a}nchez}, \citenamefont {Alonso},
  \citenamefont {Pedrosa}, \citenamefont {Hidalgo}, \citenamefont
  {De~Aguilera},\ and\ \citenamefont {Fraguas}}]{van2014use}%
  \BibitemOpen
  \bibfield  {author} {\bibinfo {author} {\bibfnamefont {B.~P.}\ \bibnamefont
  {Van~Milligen}}, \bibinfo {author} {\bibfnamefont {E.}~\bibnamefont
  {S{\'a}nchez}}, \bibinfo {author} {\bibfnamefont {A.}~\bibnamefont {Alonso}},
  \bibinfo {author} {\bibfnamefont {M.}~\bibnamefont {Pedrosa}}, \bibinfo
  {author} {\bibfnamefont {C.}~\bibnamefont {Hidalgo}}, \bibinfo {author}
  {\bibfnamefont {A.~M.}\ \bibnamefont {De~Aguilera}},\ and\ \bibinfo {author}
  {\bibfnamefont {A.~L.}\ \bibnamefont {Fraguas}},\ }\bibfield  {title}
  {\bibinfo {title} {The use of the biorthogonal decomposition for the
  identification of zonal flows at {TJ-II}},\ }\href@noop {} {\bibfield
  {journal} {\bibinfo  {journal} {Plasma Physics and Controlled Fusion}\
  }\textbf {\bibinfo {volume} {57}},\ \bibinfo {pages} {025005} (\bibinfo
  {year} {2014})}\BibitemShut {NoStop}%
\bibitem [{\citenamefont {Hansen}\ \emph {et~al.}(2015)\citenamefont {Hansen},
  \citenamefont {Victor}, \citenamefont {Morgan}, \citenamefont {Jarboe},
  \citenamefont {Hossack}, \citenamefont {Marklin}, \citenamefont {Nelson},\
  and\ \citenamefont {Sutherland}}]{hansen2015numerical}%
  \BibitemOpen
  \bibfield  {author} {\bibinfo {author} {\bibfnamefont {C.}~\bibnamefont
  {Hansen}}, \bibinfo {author} {\bibfnamefont {B.}~\bibnamefont {Victor}},
  \bibinfo {author} {\bibfnamefont {K.}~\bibnamefont {Morgan}}, \bibinfo
  {author} {\bibfnamefont {T.}~\bibnamefont {Jarboe}}, \bibinfo {author}
  {\bibfnamefont {A.}~\bibnamefont {Hossack}}, \bibinfo {author} {\bibfnamefont
  {G.}~\bibnamefont {Marklin}}, \bibinfo {author} {\bibfnamefont
  {B.}~\bibnamefont {Nelson}},\ and\ \bibinfo {author} {\bibfnamefont
  {D.}~\bibnamefont {Sutherland}},\ }\bibfield  {title} {\bibinfo {title}
  {Numerical studies and metric development for validation of
  magnetohydrodynamic models on the {HIT-SI} experiment},\ }\href@noop {}
  {\bibfield  {journal} {\bibinfo  {journal} {Physics of Plasmas}\ }\textbf
  {\bibinfo {volume} {22}},\ \bibinfo {pages} {056105} (\bibinfo {year}
  {2015})}\BibitemShut {NoStop}%
\bibitem [{\citenamefont {Schnack}\ \emph {et~al.}(2006)\citenamefont
  {Schnack}, \citenamefont {Barnes}, \citenamefont {Brennan}, \citenamefont
  {Hegna}, \citenamefont {Held}, \citenamefont {Kim}, \citenamefont {Kruger},
  \citenamefont {Pankin},\ and\ \citenamefont {Sovinec}}]{Schnack2006}%
  \BibitemOpen
  \bibfield  {author} {\bibinfo {author} {\bibfnamefont {D.~D.}\ \bibnamefont
  {Schnack}}, \bibinfo {author} {\bibfnamefont {D.~C.}\ \bibnamefont {Barnes}},
  \bibinfo {author} {\bibfnamefont {D.~P.}\ \bibnamefont {Brennan}}, \bibinfo
  {author} {\bibfnamefont {C.~C.}\ \bibnamefont {Hegna}}, \bibinfo {author}
  {\bibfnamefont {E.}~\bibnamefont {Held}}, \bibinfo {author} {\bibfnamefont
  {C.~C.}\ \bibnamefont {Kim}}, \bibinfo {author} {\bibfnamefont {S.~E.}\
  \bibnamefont {Kruger}}, \bibinfo {author} {\bibfnamefont {A.~Y.}\
  \bibnamefont {Pankin}},\ and\ \bibinfo {author} {\bibfnamefont {C.~R.}\
  \bibnamefont {Sovinec}},\ }\bibfield  {title} {\bibinfo {title}
  {Computational modeling of fully ionized magnetized plasmas using the fluid
  approximation},\ }\href@noop {} {\bibfield  {journal} {\bibinfo  {journal}
  {Physics of Plasmas}\ }\textbf {\bibinfo {volume} {13}},\ \bibinfo {pages}
  {058103} (\bibinfo {year} {2006})}\BibitemShut {NoStop}%
\bibitem [{\citenamefont {Ma}\ and\ \citenamefont {Bhattacharjee}()}]{Ma2001}%
  \BibitemOpen
  \bibfield  {author} {\bibinfo {author} {\bibfnamefont {Z.~W.}\ \bibnamefont
  {Ma}}\ and\ \bibinfo {author} {\bibfnamefont {A.}~\bibnamefont
  {Bhattacharjee}},\ }\bibfield  {title} {\bibinfo {title} {Hall
  magnetohydrodynamic reconnection: The geospace environment modeling
  challenge},\ }\href@noop {} {\bibfield  {journal} {\bibinfo  {journal}
  {Journal of Geophysical Research: Space Physics}\ }\textbf {\bibinfo {volume}
  {106}},\ \bibinfo {pages} {3773}}\BibitemShut {NoStop}%
\bibitem [{\citenamefont {Krishan}\ and\ \citenamefont
  {Mahajan}()}]{Krishan2004}%
  \BibitemOpen
  \bibfield  {author} {\bibinfo {author} {\bibfnamefont {V.}~\bibnamefont
  {Krishan}}\ and\ \bibinfo {author} {\bibfnamefont {S.~M.}\ \bibnamefont
  {Mahajan}},\ }\bibfield  {title} {\bibinfo {title} {Magnetic fluctuations and
  {H}all magnetohydrodynamic turbulence in the solar wind},\ }\href@noop {}
  {\bibfield  {journal} {\bibinfo  {journal} {Journal of Geophysical Research:
  Space Physics}\ }\textbf {\bibinfo {volume} {109}}}\BibitemShut {NoStop}%
\bibitem [{\citenamefont {Ebrahimi}\ \emph {et~al.}(2011)\citenamefont
  {Ebrahimi}, \citenamefont {Lefebvre}, \citenamefont {Forest},\ and\
  \citenamefont {Bhattacharjee}}]{Ebrahimi2011}%
  \BibitemOpen
  \bibfield  {author} {\bibinfo {author} {\bibfnamefont {F.}~\bibnamefont
  {Ebrahimi}}, \bibinfo {author} {\bibfnamefont {B.}~\bibnamefont {Lefebvre}},
  \bibinfo {author} {\bibfnamefont {C.~B.}\ \bibnamefont {Forest}},\ and\
  \bibinfo {author} {\bibfnamefont {A.}~\bibnamefont {Bhattacharjee}},\
  }\bibfield  {title} {\bibinfo {title} {Global {H}all-{MHD} simulations of
  magnetorotational instability in a plasma {C}ouette flow experiment},\
  }\href@noop {} {\bibfield  {journal} {\bibinfo  {journal} {Physics of
  Plasmas}\ }\textbf {\bibinfo {volume} {18}},\ \bibinfo {pages} {062904}
  (\bibinfo {year} {2011})}\BibitemShut {NoStop}%
\bibitem [{\citenamefont {Ferraro}(2012)}]{Ferraro2012}%
  \BibitemOpen
  \bibfield  {author} {\bibinfo {author} {\bibfnamefont {N.~M.}\ \bibnamefont
  {Ferraro}},\ }\bibfield  {title} {\bibinfo {title} {Calculations of two-fluid
  linear response to non-axisymmetric fields in tokamaks},\ }\href@noop {}
  {\bibfield  {journal} {\bibinfo  {journal} {Physics of Plasmas}\ }\textbf
  {\bibinfo {volume} {19}},\ \bibinfo {pages} {056105} (\bibinfo {year}
  {2012})}\BibitemShut {NoStop}%
\bibitem [{\citenamefont {Kaptanoglu}\ \emph
  {et~al.}(2020{\natexlab{c}})\citenamefont {Kaptanoglu}, \citenamefont
  {Benedett}, \citenamefont {Morgan}, \citenamefont {Hansen},\ and\
  \citenamefont {Jarboe}}]{kaptanoglu2020two}%
  \BibitemOpen
  \bibfield  {author} {\bibinfo {author} {\bibfnamefont {A.~A.}\ \bibnamefont
  {Kaptanoglu}}, \bibinfo {author} {\bibfnamefont {T.~E.}\ \bibnamefont
  {Benedett}}, \bibinfo {author} {\bibfnamefont {K.~D.}\ \bibnamefont
  {Morgan}}, \bibinfo {author} {\bibfnamefont {C.~J.}\ \bibnamefont {Hansen}},\
  and\ \bibinfo {author} {\bibfnamefont {T.~R.}\ \bibnamefont {Jarboe}},\
  }\bibfield  {title} {\bibinfo {title} {Two-temperature effects in {Hall-MHD}
  simulations of the {HIT-SI} experiment},\ }\href@noop {} {\bibfield
  {journal} {\bibinfo  {journal} {Physics of Plasmas}\ }\textbf {\bibinfo
  {volume} {27}},\ \bibinfo {pages} {072505} (\bibinfo {year}
  {2020}{\natexlab{c}})}\BibitemShut {NoStop}%
\bibitem [{\citenamefont {Brunton}\ and\ \citenamefont
  {Kutz}(2019)}]{brunton2019data}%
  \BibitemOpen
  \bibfield  {author} {\bibinfo {author} {\bibfnamefont {S.~L.}\ \bibnamefont
  {Brunton}}\ and\ \bibinfo {author} {\bibfnamefont {J.~N.}\ \bibnamefont
  {Kutz}},\ }\href@noop {} {\emph {\bibinfo {title} {Data-driven science and
  engineering: Machine learning, dynamical systems, and control}}}\ (\bibinfo
  {publisher} {Cambridge University Press},\ \bibinfo {year}
  {2019})\BibitemShut {NoStop}%
\bibitem [{\citenamefont {Golub}\ \emph {et~al.}(1996)\citenamefont {Golub}
  \emph {et~al.}}]{golub1996cf}%
  \BibitemOpen
  \bibfield  {author} {\bibinfo {author} {\bibfnamefont {G.~H.}\ \bibnamefont
  {Golub}} \emph {et~al.},\ }\bibfield  {title} {\bibinfo {title} {Matrix
  computations},\ }\href@noop {} {\bibfield  {journal} {\bibinfo  {journal}
  {The Johns Hopkins}\ } (\bibinfo {year} {1996})}\BibitemShut {NoStop}%
\bibitem [{\citenamefont {Frieze}\ \emph {et~al.}(2004)\citenamefont {Frieze},
  \citenamefont {Kannan},\ and\ \citenamefont {Vempala}}]{frieze2004fast}%
  \BibitemOpen
  \bibfield  {author} {\bibinfo {author} {\bibfnamefont {A.}~\bibnamefont
  {Frieze}}, \bibinfo {author} {\bibfnamefont {R.}~\bibnamefont {Kannan}},\
  and\ \bibinfo {author} {\bibfnamefont {S.}~\bibnamefont {Vempala}},\
  }\bibfield  {title} {\bibinfo {title} {Fast {M}onte-{C}arlo algorithms for
  finding low-rank approximations},\ }\href@noop {} {\bibfield  {journal}
  {\bibinfo  {journal} {Journal of the ACM (JACM)}\ }\textbf {\bibinfo {volume}
  {51}},\ \bibinfo {pages} {1025} (\bibinfo {year} {2004})}\BibitemShut
  {NoStop}%
\bibitem [{\citenamefont {Liberty}\ \emph {et~al.}(2007)\citenamefont
  {Liberty}, \citenamefont {Woolfe}, \citenamefont {Martinsson}, \citenamefont
  {Rokhlin},\ and\ \citenamefont {Tygert}}]{liberty2007randomized}%
  \BibitemOpen
  \bibfield  {author} {\bibinfo {author} {\bibfnamefont {E.}~\bibnamefont
  {Liberty}}, \bibinfo {author} {\bibfnamefont {F.}~\bibnamefont {Woolfe}},
  \bibinfo {author} {\bibfnamefont {P.-G.}\ \bibnamefont {Martinsson}},
  \bibinfo {author} {\bibfnamefont {V.}~\bibnamefont {Rokhlin}},\ and\ \bibinfo
  {author} {\bibfnamefont {M.}~\bibnamefont {Tygert}},\ }\bibfield  {title}
  {\bibinfo {title} {Randomized algorithms for the low-rank approximation of
  matrices},\ }\href@noop {} {\bibfield  {journal} {\bibinfo  {journal}
  {Proceedings of the National Academy of Sciences}\ }\textbf {\bibinfo
  {volume} {104}},\ \bibinfo {pages} {20167} (\bibinfo {year}
  {2007})}\BibitemShut {NoStop}%
\bibitem [{\citenamefont {Woolfe}\ \emph {et~al.}(2008)\citenamefont {Woolfe},
  \citenamefont {Liberty}, \citenamefont {Rokhlin},\ and\ \citenamefont
  {Tygert}}]{woolfe2008fast}%
  \BibitemOpen
  \bibfield  {author} {\bibinfo {author} {\bibfnamefont {F.}~\bibnamefont
  {Woolfe}}, \bibinfo {author} {\bibfnamefont {E.}~\bibnamefont {Liberty}},
  \bibinfo {author} {\bibfnamefont {V.}~\bibnamefont {Rokhlin}},\ and\ \bibinfo
  {author} {\bibfnamefont {M.}~\bibnamefont {Tygert}},\ }\bibfield  {title}
  {\bibinfo {title} {A fast randomized algorithm for the approximation of
  matrices},\ }\href@noop {} {\bibfield  {journal} {\bibinfo  {journal}
  {Applied and Computational Harmonic Analysis}\ }\textbf {\bibinfo {volume}
  {25}},\ \bibinfo {pages} {335} (\bibinfo {year} {2008})}\BibitemShut
  {NoStop}%
\bibitem [{\citenamefont {Jarboe}\ \emph {et~al.}(2006)\citenamefont {Jarboe},
  \citenamefont {Hamp}, \citenamefont {Marklin}, \citenamefont {Nelson},
  \citenamefont {O’Neill}, \citenamefont {Redd}, \citenamefont {Sieck},
  \citenamefont {Smith},\ and\ \citenamefont {Wrobel}}]{jarboe2006spheromak}%
  \BibitemOpen
  \bibfield  {author} {\bibinfo {author} {\bibfnamefont {T.}~\bibnamefont
  {Jarboe}}, \bibinfo {author} {\bibfnamefont {W.}~\bibnamefont {Hamp}},
  \bibinfo {author} {\bibfnamefont {G.}~\bibnamefont {Marklin}}, \bibinfo
  {author} {\bibfnamefont {B.}~\bibnamefont {Nelson}}, \bibinfo {author}
  {\bibfnamefont {R.}~\bibnamefont {O’Neill}}, \bibinfo {author}
  {\bibfnamefont {A.}~\bibnamefont {Redd}}, \bibinfo {author} {\bibfnamefont
  {P.}~\bibnamefont {Sieck}}, \bibinfo {author} {\bibfnamefont
  {R.}~\bibnamefont {Smith}},\ and\ \bibinfo {author} {\bibfnamefont
  {J.}~\bibnamefont {Wrobel}},\ }\bibfield  {title} {\bibinfo {title}
  {Spheromak formation by steady inductive helicity injection},\ }\href@noop {}
  {\bibfield  {journal} {\bibinfo  {journal} {Physical review letters}\
  }\textbf {\bibinfo {volume} {97}},\ \bibinfo {pages} {115003} (\bibinfo
  {year} {2006})}\BibitemShut {NoStop}%
\bibitem [{\citenamefont {Galtier}(2016)}]{galtier2016introduction}%
  \BibitemOpen
  \bibfield  {author} {\bibinfo {author} {\bibfnamefont {S.}~\bibnamefont
  {Galtier}},\ }\href@noop {} {\emph {\bibinfo {title} {Introduction to modern
  magnetohydrodynamics}}}\ (\bibinfo  {publisher} {Cambridge University
  Press},\ \bibinfo {year} {2016})\BibitemShut {NoStop}%
\bibitem [{\citenamefont {Loiseau}\ \emph {et~al.}(2018)\citenamefont
  {Loiseau}, \citenamefont {Noack},\ and\ \citenamefont
  {Brunton}}]{Loiseau2018jfm}%
  \BibitemOpen
  \bibfield  {author} {\bibinfo {author} {\bibfnamefont {J.-C.}\ \bibnamefont
  {Loiseau}}, \bibinfo {author} {\bibfnamefont {B.~R.}\ \bibnamefont {Noack}},\
  and\ \bibinfo {author} {\bibfnamefont {S.~L.}\ \bibnamefont {Brunton}},\
  }\bibfield  {title} {\bibinfo {title} {Sparse reduced-order modeling:
  sensor-based dynamics to full-state estimation},\ }\href@noop {} {\bibfield
  {journal} {\bibinfo  {journal} {Journal of Fluid Mechanics}\ }\textbf
  {\bibinfo {volume} {844}},\ \bibinfo {pages} {459} (\bibinfo {year}
  {2018})}\BibitemShut {NoStop}%
\bibitem [{\citenamefont {Benner}\ \emph {et~al.}(2015)\citenamefont {Benner},
  \citenamefont {Gugercin},\ and\ \citenamefont {Willcox}}]{benner2015survey}%
  \BibitemOpen
  \bibfield  {author} {\bibinfo {author} {\bibfnamefont {P.}~\bibnamefont
  {Benner}}, \bibinfo {author} {\bibfnamefont {S.}~\bibnamefont {Gugercin}},\
  and\ \bibinfo {author} {\bibfnamefont {K.}~\bibnamefont {Willcox}},\
  }\bibfield  {title} {\bibinfo {title} {A survey of projection-based model
  reduction methods for parametric dynamical systems},\ }\href@noop {}
  {\bibfield  {journal} {\bibinfo  {journal} {SIAM review}\ }\textbf {\bibinfo
  {volume} {57}},\ \bibinfo {pages} {483} (\bibinfo {year} {2015})}\BibitemShut
  {NoStop}%
\bibitem [{\citenamefont {Chaturantabut}\ and\ \citenamefont
  {Sorensen}(2009)}]{chaturantabut2009discrete}%
  \BibitemOpen
  \bibfield  {author} {\bibinfo {author} {\bibfnamefont {S.}~\bibnamefont
  {Chaturantabut}}\ and\ \bibinfo {author} {\bibfnamefont {D.~C.}\ \bibnamefont
  {Sorensen}},\ }\bibfield  {title} {\bibinfo {title} {Discrete empirical
  interpolation for nonlinear model reduction},\ }in\ \href@noop {} {\emph
  {\bibinfo {booktitle} {Proceedings of the 48th IEEE Conference on Decision
  and Control (CDC) held jointly with 2009 28th Chinese Control Conference}}}\
  (\bibinfo {organization} {IEEE},\ \bibinfo {year} {2009})\ pp.\ \bibinfo
  {pages} {4316--4321}\BibitemShut {NoStop}%
\bibitem [{\citenamefont {Drmac}\ and\ \citenamefont
  {Gugercin}(2016)}]{drmac2016new}%
  \BibitemOpen
  \bibfield  {author} {\bibinfo {author} {\bibfnamefont {Z.}~\bibnamefont
  {Drmac}}\ and\ \bibinfo {author} {\bibfnamefont {S.}~\bibnamefont
  {Gugercin}},\ }\bibfield  {title} {\bibinfo {title} {A new selection operator
  for the discrete empirical interpolation method---improved a priori error
  bound and extensions},\ }\href@noop {} {\bibfield  {journal} {\bibinfo
  {journal} {SIAM Journal on Scientific Computing}\ }\textbf {\bibinfo {volume}
  {38}},\ \bibinfo {pages} {A631} (\bibinfo {year} {2016})}\BibitemShut
  {NoStop}%
\bibitem [{\citenamefont {Astrid}\ \emph {et~al.}(2008)\citenamefont {Astrid},
  \citenamefont {Weiland}, \citenamefont {Willcox},\ and\ \citenamefont
  {Backx}}]{astrid2008missing}%
  \BibitemOpen
  \bibfield  {author} {\bibinfo {author} {\bibfnamefont {P.}~\bibnamefont
  {Astrid}}, \bibinfo {author} {\bibfnamefont {S.}~\bibnamefont {Weiland}},
  \bibinfo {author} {\bibfnamefont {K.}~\bibnamefont {Willcox}},\ and\ \bibinfo
  {author} {\bibfnamefont {T.}~\bibnamefont {Backx}},\ }\bibfield  {title}
  {\bibinfo {title} {Missing point estimation in models described by proper
  orthogonal decomposition},\ }\href@noop {} {\bibfield  {journal} {\bibinfo
  {journal} {IEEE Transactions on Automatic Control}\ }\textbf {\bibinfo
  {volume} {53}},\ \bibinfo {pages} {2237} (\bibinfo {year}
  {2008})}\BibitemShut {NoStop}%
\bibitem [{\citenamefont {Willcox}(2006)}]{willcox2006unsteady}%
  \BibitemOpen
  \bibfield  {author} {\bibinfo {author} {\bibfnamefont {K.}~\bibnamefont
  {Willcox}},\ }\bibfield  {title} {\bibinfo {title} {Unsteady flow sensing and
  estimation via the gappy proper orthogonal decomposition},\ }\href@noop {}
  {\bibfield  {journal} {\bibinfo  {journal} {Computers \& Fluids}\ }\textbf
  {\bibinfo {volume} {35}},\ \bibinfo {pages} {208} (\bibinfo {year}
  {2006})}\BibitemShut {NoStop}%
\bibitem [{\citenamefont {Carlberg}\ \emph {et~al.}(2013)\citenamefont
  {Carlberg}, \citenamefont {Farhat}, \citenamefont {Cortial},\ and\
  \citenamefont {Amsallem}}]{carlberg2013gnat}%
  \BibitemOpen
  \bibfield  {author} {\bibinfo {author} {\bibfnamefont {K.}~\bibnamefont
  {Carlberg}}, \bibinfo {author} {\bibfnamefont {C.}~\bibnamefont {Farhat}},
  \bibinfo {author} {\bibfnamefont {J.}~\bibnamefont {Cortial}},\ and\ \bibinfo
  {author} {\bibfnamefont {D.}~\bibnamefont {Amsallem}},\ }\bibfield  {title}
  {\bibinfo {title} {The {GNAT} method for nonlinear model reduction: effective
  implementation and application to computational fluid dynamics and turbulent
  flows},\ }\href@noop {} {\bibfield  {journal} {\bibinfo  {journal} {Journal
  of Computational Physics}\ }\textbf {\bibinfo {volume} {242}},\ \bibinfo
  {pages} {623} (\bibinfo {year} {2013})}\BibitemShut {NoStop}%
\bibitem [{\citenamefont {Brunton}\ \emph {et~al.}(2016)\citenamefont
  {Brunton}, \citenamefont {Proctor},\ and\ \citenamefont
  {Kutz}}]{Brunton2016pnas}%
  \BibitemOpen
  \bibfield  {author} {\bibinfo {author} {\bibfnamefont {S.~L.}\ \bibnamefont
  {Brunton}}, \bibinfo {author} {\bibfnamefont {J.~L.}\ \bibnamefont
  {Proctor}},\ and\ \bibinfo {author} {\bibfnamefont {J.~N.}\ \bibnamefont
  {Kutz}},\ }\bibfield  {title} {\bibinfo {title} {Discovering governing
  equations from data by sparse identification of nonlinear dynamical
  systems},\ }\href@noop {} {\bibfield  {journal} {\bibinfo  {journal}
  {Proceedings of the National Academy of Sciences}\ }\textbf {\bibinfo
  {volume} {113}},\ \bibinfo {pages} {3932} (\bibinfo {year}
  {2016})}\BibitemShut {NoStop}%
\bibitem [{\citenamefont {Morrison}\ and\ \citenamefont
  {Greene}(1980)}]{morrison1980noncanonical}%
  \BibitemOpen
  \bibfield  {author} {\bibinfo {author} {\bibfnamefont {P.~J.}\ \bibnamefont
  {Morrison}}\ and\ \bibinfo {author} {\bibfnamefont {J.~M.}\ \bibnamefont
  {Greene}},\ }\bibfield  {title} {\bibinfo {title} {Noncanonical hamiltonian
  density formulation of hydrodynamics and ideal magnetohydrodynamics},\
  }\href@noop {} {\bibfield  {journal} {\bibinfo  {journal} {Physical Review
  Letters}\ }\textbf {\bibinfo {volume} {45}},\ \bibinfo {pages} {790}
  (\bibinfo {year} {1980})}\BibitemShut {NoStop}%
\bibitem [{\citenamefont {Yoshida}\ and\ \citenamefont
  {Hameiri}(2013)}]{yoshida2013canonical}%
  \BibitemOpen
  \bibfield  {author} {\bibinfo {author} {\bibfnamefont {Z.}~\bibnamefont
  {Yoshida}}\ and\ \bibinfo {author} {\bibfnamefont {E.}~\bibnamefont
  {Hameiri}},\ }\bibfield  {title} {\bibinfo {title} {Canonical hamiltonian
  mechanics of hall magnetohydrodynamics and its limit to ideal
  magnetohydrodynamics},\ }\href@noop {} {\bibfield  {journal} {\bibinfo
  {journal} {Journal of Physics A: Mathematical and Theoretical}\ }\textbf
  {\bibinfo {volume} {46}},\ \bibinfo {pages} {335502} (\bibinfo {year}
  {2013})}\BibitemShut {NoStop}%
\bibitem [{\citenamefont {Abdelhamid}\ \emph {et~al.}(2015)\citenamefont
  {Abdelhamid}, \citenamefont {Kawazura},\ and\ \citenamefont
  {Yoshida}}]{abdelhamid2015hamiltonian}%
  \BibitemOpen
  \bibfield  {author} {\bibinfo {author} {\bibfnamefont {H.~M.}\ \bibnamefont
  {Abdelhamid}}, \bibinfo {author} {\bibfnamefont {Y.}~\bibnamefont
  {Kawazura}},\ and\ \bibinfo {author} {\bibfnamefont {Z.}~\bibnamefont
  {Yoshida}},\ }\bibfield  {title} {\bibinfo {title} {Hamiltonian formalism of
  extended magnetohydrodynamics},\ }\href@noop {} {\bibfield  {journal}
  {\bibinfo  {journal} {Journal of Physics A: Mathematical and Theoretical}\
  }\textbf {\bibinfo {volume} {48}},\ \bibinfo {pages} {235502} (\bibinfo
  {year} {2015})}\BibitemShut {NoStop}%
\bibitem [{\citenamefont {Chu}\ and\ \citenamefont
  {Hayashibe}(2020)}]{chu2020discovering}%
  \BibitemOpen
  \bibfield  {author} {\bibinfo {author} {\bibfnamefont {H.~K.}\ \bibnamefont
  {Chu}}\ and\ \bibinfo {author} {\bibfnamefont {M.}~\bibnamefont
  {Hayashibe}},\ }\bibfield  {title} {\bibinfo {title} {Discovering
  interpretable dynamics by sparsity promotion on energy and the lagrangian},\
  }\href@noop {} {\bibfield  {journal} {\bibinfo  {journal} {IEEE Robotics and
  Automation Letters}\ }\textbf {\bibinfo {volume} {5}},\ \bibinfo {pages}
  {2154} (\bibinfo {year} {2020})}\BibitemShut {NoStop}%
\end{thebibliography}%

\end{document}